\documentclass[amsmath,prb,twocolumn,showpacs]{revtex4}

\usepackage{color}
\usepackage{amsfonts}
\usepackage{graphicx}

\begin{document}

\title{A non-equilibrium equation-of-motion approach to quantum transport utilizing projection operators}
\author{Maicol A. Ochoa}
\email{mochoadaza@ucsd.edu}
\affiliation{Department of Chemistry \& Biochemistry, University of California San Diego, La Jolla, CA 92093, USA}
\author{Michael Galperin}
\email{migalperin@ucsd.edu}
\affiliation{Department of Chemistry \& Biochemistry, University of California San Diego, La Jolla, CA 92093, USA}
\author{Mark A. Ratner}
\email{ratner@northwestern.edu}
\affiliation{Department of Chemistry, Northwestern University, Evanston, IL 60208, USA}

\begin{abstract}
We consider a projection operator approach to the non-equilbrium Green function 
equation-of-motion (PO-NEGF EOM) method. 
The technique resolves problems of arbitrariness 
in truncation  of an infinite chain of EOMs, and prevents violation of symmetry 
relations resulting from the truncation. The approach, originally developed by
Tserkovnikov [Theor. Math. Phys. \textbf{118}, 85 (1999)] for equilibrium systems, 
is reformulated to be applicable to time-dependent non-equilibrium situations.
We derive a canonical form of EOMs, thus explicitly demonstrating
a proper result for the non-equilibrium atomic limit in junction problems.
A simple practical scheme applicable to quantum transport simulations is 
formulated. We perform numerical simulations within simple models,
and compare results of the approach to other techniques, and (where available) also 
to exact results. 
\end{abstract}

\pacs{85.65.+h 73.23.-b 73.63.Kv 85.35.-p}

\maketitle
\section{Introduction}\label{intro}
Since the first theoretical prediction of the possibility to utilize
molecules as active elements in junctions,\cite{AviramRatnerCPL74}
molecular electronics has been driven by developments in experimental
abilities to fabricate and control molecular nanostructures.
Inelastic electron tunneling spectroscopy (IETS),\cite{GalperinRatnerNitzanTroisiScience08,GalperinRatnerNitzanJPCM07} 
molecular optoelectronics,\cite{GalperinNitzanPCCP12}
molecular nanoplasmonics,\cite{SchatzRatnerRPP12} 
and molecular spintronics\cite{SanvitoJCTN06,WernsdorferNM08,WernsdorferIJN10,SanvitoCSR11,NaamanWaldeckJPCL12} are 
branches of molecular electronics, where experimental data
require development of an adequate theoretical techniques.  

Resonant IETS with its strong electron-vibration coupling and anharmonic
effects,\cite{McEuenNature00}
as well as observation of breakdown of the Born-Oppenheimer 
approximation,\cite{ReppNatPhys10} requires formulation in the basis of 
vibronic states.\cite{BoncaTrugmanPRL95,WhiteGalperinPCCP12}
Similarly, in optoelectronic devices, where covalent bonding
between molecule and contacts results in hybridization between
molecular excitation and plasmon resonances,\cite{NordlanderNatCommun12}
separation into electronic and plasmonic degrees of freedom is 
inadequate.\cite{WhiteFainbergGalperinJPCL12}
Finally, physics of single molecule magnet devices is achieved 
most conveniently by diagonalizing the multi-spin 
Hamiltonian.\cite{RalphNL06,HendricksonPRL11}

Development of theoretical methods capable of describing transport
in molecular junctions in the language of many-body states of an isolated
system - the nonequilibrium atomic limit - is necessary for
adequate description in the situations similar to those mentioned above.\cite{WhiteOchoaGalperinJPCC14}
Examples of such formulations include scattering theory 
methods,\cite{BoncaTrugmanPRL95,HauleBoncaPRB99,EmberlyKirczenowPRB00,SeidemanAccChemRes10,GalperinNitzanJPCB13}
quantum master equation approaches,\cite{KochVonOppenPRL05,KochRaikhVonOppenPRL06,GhoshDattaPRB07,LeijnseWegewijsPRB08,EspGalpPRB09,EspGalpJPCC10,MayNL12,WegewijsPRB13} 
pseudoparticle\cite{EcksteinWernerPRB10,OhAhnBubanjaPRB11,WhiteGalperinPCCP12,WhiteFainbergGalperinJPCL12,BronoldFehskePRB12,WhiteNitzanJCP13,WhiteTretiakNL14} and 
Hubbard\cite{SandalovIJQC03,FranssonPRB05,SandalovPRB07,GalperinNitzanRatnerPRB08,Yeganeh2009} non-equilibrium Green function techniques.

A simple general alternative is the equation-of-motion (EOM) 
method\cite{BonchBruevich_Tyablikov_1962,KadanoffBaym_1962} formulated 
for transport problems on the Keldysh contour.\cite{BulkaKostyrkoPRB04,KostyrkoBulkaPRB05,SpickaVelikyKarlovaPhysE05,GalperinNitzanRatnerPRB06,GalperinNitzanRatnerPRB07,TrochaBarnasJPCM08,SwirkowiczBarnasWilczynskyJMMM09,FainbergPRB11,LevyRabaniJCP13,LevyRabaniJPCM13}
EOM permits working with correlation functions of any operators to produce
(in general) an infinite chain of equations of motion. 
The main drawback of the method is the necessity to make an uncontrolled 
approximation to close the chain of equations. Usually, a justification
for such an approximation can be obtained only a posteriori. 
Moreover, uncontrolled decoupling of correlation functions may result 
in complications related to loss of proper commutation relations
between the decoupled operators.\cite{BonchBruevich_Tyablikov_1962}
Violation of symmetry relations due to such 
an uncontrolled decoupling in quantum transport problems was discussed in 
Refs.~\onlinecite{LevyRabaniJCP13,LevyRabaniJPCM13}.

Projection operators are often employed to derive an exact 
quantum master equation (QME) for a system coupled to 
baths.\cite{Zwanzig_2001,BreuerPetruccione_2003,Nitzan_2006,vanKampen_2007,VanVlietJPhysChemSol88,KoideProgTheorPhys02,WeimerEPJST07,BreuerEPJST07,SchmidtEPJB09}
In the theory of equilibrium Green functions, projection techniques are
sometimes utilized to close the chain of EOMs.\cite{RothPRL68,IchiyanagiJPhysSocJpn72,GoryachevKuzminOvchinnikovJPhysC82,RuckensteinIntJModPhysB89}
In particular, an approach developed by Tserkovnikov\cite{TserkovnikovTheorMathPhys81,TserkovnikovTheorMathPhys99}
guarantees that truncating an infinite chain of equations at step $n$
includes correlations between the originally chosen set of operators
to order $n$. The approach was shown to be equivalent to Mori's 
method\cite{MoriProgTheorPhys65,RyuChoiPRB91,OkadaSawadaKurodaJPhysSocJpn95}
employed in QME derivations. 

Here we formulate the approach of Refs.~\onlinecite{TserkovnikovTheorMathPhys81,TserkovnikovTheorMathPhys99,PlakidaTheorMathPhys11} 
on the Keldysh contour, and discuss its applicability to transport problems.
In particular, we demonstrate that the approach  permits a formulation of 
a symmetrized Dyson-type equation which by construction resolves 
symmetry violations discussed in Refs.~\onlinecite{LevyRabaniJCP13,LevyRabaniJPCM13}.
Also we argue that the approach resolves the problems
of the first Hubbard (HIA) approximation related to Hermiticity of a resulting reduced 
density matrix.\cite{GalperinNitzanRatnerPRB08,PedersenPRB09} 

After introducing the generalization of the scheme in Section~\ref{method}A,
in Section~\ref{method}B we derive a Dyson-type EOM for effective canonical 
quasi-particles, which is crucial for construction of a proper diagrammatic 
expansion.\cite{ShastryPRB10} 
Then in Section~\ref{method}C we discuss application of the resulting scheme to transport
in junctions. Section~\ref{numres} presents numerical results.
Section~\ref{conclusions} concludes.

\section{Method}\label{method}
We start by generalizing the scheme of Refs.~\onlinecite{TserkovnikovTheorMathPhys81,TserkovnikovTheorMathPhys99,PlakidaTheorMathPhys11} to the realm of 
open quantum systems far from equilibrium in Section~\ref{methodA}. 
General consideration follows by derivation of the EOM for canonical Green
function in Section~\ref{methodB}. Finally, in Section~\ref{methodC}
we discuss application of the scheme to transport in junctions.
\subsection{EOM for irreducible Green functions}\label{methodA}
Following Ref.~\onlinecite{TserkovnikovTheorMathPhys99} we start
by introducing {\em a column vector operator}\/ $\hat a_1$.
Both choice and number $M_1$ of its elements depend on 
the particular problem (see Section~\ref{numres} for examples).
This flexibility of choice is an advantage, allowing to greatly simplify solution 
 once the initial vector-operator is taken according to physics of a problem 
 (see examples in Section~\ref{numres}). The moment the original operator has been 
 chosen the methodology provides a tool for taking account of correlations 
 between the operators in an ordered fashion.

EOMs (in the Heisenberg picture) for this vector-operator and 
its Hermitian conjugate are (here and below $\hbar =1$)
\begin{align}
\label{EOMan}
i\frac{\partial}{\partial\tau}\hat a_n(\tau) \equiv&
\left[\hat a_n;\hat H\right](\tau) 
\\ &=
\mathbf{\varepsilon}_n(\tau) \hat a_n(\tau) 
+ \mathbf{\nu}_{n,n+1}(\tau)\hat a_{n+1}(\tau),
\nonumber 
\end{align}
\begin{align}
\label{EOMand}
-i\frac{\partial}{\partial\tau'}\hat a_n^\dagger(\tau') \equiv&
\left[\hat H;\hat a_n^\dagger\right](\tau') 
\\ &=
\hat a_n^\dagger(\tau')\bar{\mathbf{\varepsilon}}_n(\tau') + 
\hat a_{n+1}^\dagger(\tau') \mathbf{\nu}_{n+1,n}(\tau').
\nonumber
\end{align}
Here $n=1,2,3,\ldots$ is the index for the infinite sequence of operators
defined by the EOMs (\ref{EOMan}) and (\ref{EOMand}),
$\tau$ and $\tau'$ are contour variables, and $\hat H$ is the total 
Hamiltonian. $\mathbf{\varepsilon}_n$ and $\mathbf{\nu}_{n,n+1}$
are matrices of size $M_n\times M_n$ and $M_n\times M_{n+1}$, respectively, 
defined by commutation of operators constituting vector the $\hat a_n$ 
with the Hamiltonian $\hat H$.
$\bar{\mathbf{\varepsilon}}_n=\mathbf{\varepsilon}_n^\dagger$
and $\mathbf{\nu}_{n+1,n}=\mathbf{\nu}_{n,n+1}^\dagger$.
Note that $\hat a_n^\dagger$ is {\em a row vector operator} with elements
being Hermitian conjugates of those in $\hat a_n$.
Here and below $\mathbf{\varepsilon}_n(\tau)$ indicates values of the
matrix at physical time $t$ corresponding to contour variable $\tau$
(and similar for other matrices). Such dependence appears when
the Hamiltonian contains a time-dependent process.

The scalar product $\mathbf{P}_{a,b}$ of two arbitrary vector operators 
$\hat a$ and $\hat b$
is defined as an average over a non-equlibrium state of a commutator
(if at least one of the vectors is of Bose type) or an anti-commutator
(if both operators are of Fermi type)
\begin{equation}
\label{defP}
 \mathbf{P}_{a,b} \equiv \left\langle\left[\hat a;\hat b^\dagger\right]_{\pm}\right\rangle.
\end{equation}
Note that $\mathbf{P}_{a,b}$ is a matrix of size $M_a\times M_b$.

Utilizing this definition of scalar product, Eq.(\ref{defP}), we next introduce
{\em orthogonalized vector operators}
\begin{align}
 \label{defAn}
 \hat A_n \equiv& \hat a_n - \sum_{i=1}^{n-1}
 \mathbf{P}_{a_n,A_i}\,\mathbf{P}_{A_i,A_i}^{-1}\,\hat A_i 
 \\
 \label{defAnd}
 \hat A_n^\dagger \equiv& \hat a_n^\dagger - \sum_{i=1}^{n-1}
 \hat A_i^\dagger\,\mathbf{P}_{A_i,A_i}^{-1}\,\mathbf{P}_{A_i,a_n}
\end{align}
where $\hat A_1\equiv\hat a_1$.
Here and below we use capital letters for orthogonalized vector operators.
Note that in Eqs.~(\ref{defAn}) and (\ref{defAnd}), as well as below, we assume 
existence of inverse of the spectral weight matrix, Eq.(\ref{defP}).
While this property cannot be guaranteed in general, it can be proven for 
Fermi type excitations in a system held at finite temperature (see Appendix~\ref{app1}).
These are the conditions of main interest for quantum transport problems in junctions,
and this is the situation we consider in the paper. 

Our goal is to evaluate the non-equilibrium Green function of the 
operators of interest $\hat a_1$
\begin{equation}
 \label{defG}
 \mathbf{G}_{a_1,a_1}(\tau,\tau') \equiv -i\left\langle
 T_c\,\hat a_1(\tau)\,\hat a_1^\dagger(\tau')\right\rangle ,
\end{equation}
where $T_c$ is the contour ordering operator and $\mathbf{G}$
is $M_1\times M_1$ matrix. 
For future reference it is convenient to introduce 
{\em irreducible Green functions}
\begin{align}
 \label{defirrG}
& \mathbf{G}^{(n)}_{a,b}(\tau,\tau') \equiv
 \mathbf{G}^{(n-1)}_{a,b}(\tau,\tau') 
 \\ &
 - \int_c d\tau_1\int_cd\tau_2
 \mathbf{G}^{(n-1)}_{a,A_n}(\tau,\tau_1)
 \hat{\mathbf{G}}^{-1\,(n-1)}_{A_n,A_n}(\tau_1,\tau_2)
 \mathbf{G}^{(n-1)}_{A_n,b}(\tau_2,\tau')
 \nonumber
\end{align}
Here integration is over the Keldysh contour,
$\mathbf{G}^{(0)}_{a,b}(\tau,\tau')\equiv \mathbf{G}_{a,b}(\tau,\tau')$,
and $\hat{\mathbf{G}}^{-1\,(n-1)}_{A_n,A_n}(\tau_1,\tau_2)$ is the inverse operator
for the irreducible Green function 
$\mathbf{G}^{(n-1)}_{A_n,A_n}(\tau_1,\tau_2)$
\begin{align}
 \label{definvop}
 &\int_c d\tau_1 \hat{\mathbf{G}}^{-1\,(n-1)}_{A_n,A_n}(\tau,\tau_1)\,
 \mathbf{G}^{(n-1)}_{A_n,A_n}(\tau_1,\tau') =
 \\ &
 \int_c d\tau_1 \mathbf{G}^{(n-1)}_{A_n,A_n}(\tau,\tau_1)\,
 \hat{\mathbf{G}}^{-1\,(n-1)}_{A_n,A_n}(\tau_1,\tau') =
 \delta(\tau,\tau')\,\mathbf{I}_n
 \nonumber
\end{align}
where $\mathbf{I}_n$ is the $M_n\times M_n$ unit matrix.
Considering Eq.(\ref{defG}) as a scalar product in an extended Hilbert space (the one including the contour variable as part of the index), definition of irreducible Green functions in Eq.(\ref{defirrG}) is 
similar to projection in Eqs.~(\ref{defAn}) and (\ref{defAnd}), where each next generation of correlation functions
is orthogonalized relative to the previous one. This definition provides properties 
which eventually lead to formulation of a canonical Dyson-type equation
(see Eq.(\ref{EOMcqpG}) below). Note that this result would be impossible without 
the orthogonalization introduced in Eq.(\ref{defirrG}).

In terms of the irreducible Green functions, Eq.(\ref{defirrG}),
the chain of (left and right) EOMs for the Green function (\ref{defG}) is
(see Appendix~\ref{appA} for derivation)
\begin{align}
 \label{EOMirrGleft}
&\left[i\frac{\overset{\rightarrow}{\partial}}{\partial\tau}-\mathbf{\omega}_n(\tau)\right]
\mathbf{G}^{(n-1)}_{A_n,A_n}(\tau,\tau') =
 \delta(\tau,\tau')\,\mathbf{P}_{A_n,A_n}(\tau) 
 \nonumber \\ &\quad
 +
 \int_c d\tau_1\, \mathbf{P}_{A_n,A_n}(\tau)\,
 \mathbf{\Sigma}^{(n-1)}_{A_n,A_n}(\tau,\tau_1)\,
 \mathbf{G}^{(n-1)}_{A_n,A_n}(\tau_1,\tau')
\\
 \label{EOMirrGright}
&\mathbf{G}^{(n-1)}_{A_n,A_n}(\tau,\tau')
 \left[-i\frac{\overset{\leftarrow}{\partial}}{\partial\tau'}
 -\bar{\mathbf{\omega}}_n(\tau')\right] =
 \delta(\tau,\tau')\,\mathbf{P}_{A_n,A_n}(\tau) 
  \nonumber \\ &\quad
 +
 \int_c d\tau_1\, \mathbf{G}^{(n-1)}_{A_n,A_n}(\tau,\tau_1)\,
 \mathbf{\Sigma}^{(n-1)}_{A_n,A_n}(\tau_1,\tau')\,
 \mathbf{P}_{A_n,A_n}(\tau')
\end{align}
where 
\begin{align}
 \label{defomega}
 \mathbf{\omega}_n(\tau) \equiv& \mathbf{\varepsilon}_n(\tau)
 + \mathbf{\nu}_{n,n+1}(\tau)\,\mathbf{P}_{a_{n+1},A_n}(\tau)\,
   \mathbf{P}_{A_n,A_n}^{-1}(\tau)
 \nonumber \\ &
 - \mathbf{P}_{a_n,A_{n-1}}(\tau)\,\mathbf{P}_{A_{n-1},A_{n-1}}^{-1}(\tau)\,
   \mathbf{\nu}_{n-1,n}(\tau)  
\end{align}
is a normalized free evolution matrix, 
$\bar{\mathbf{\omega}}_n(\tau)\equiv\mathbf{\omega}_n^\dagger(\tau)$,
and
\begin{align}
 \label{defS}
 \mathbf{\Sigma}^{(n-1)}_{A_n,A_n}(\tau,\tau') \equiv&
 \mathbf{P}_{A_n,A_n}^{-1}(\tau)\, \mathbf{\nu}_{n,n+1}(\tau)\, 
 \mathbf{G}_{A_{n+1},A_{n+1}}^{(n)}(\tau,\tau')
 \nonumber \\ &\times
 \mathbf{\nu}_{n+1,n}(\tau')\, \mathbf{P}_{A_n,A_n}^{-1}(\tau')
\end{align}
is the irreducible self-energy of the Green function 
$\mathbf{G}_{A_{n},A_{n}}^{(n-1)}(\tau,\tau')$
due to higher order correlations.
Eqs.~(\ref{EOMirrGleft}) and (\ref{EOMirrGright}) complete 
formulation of the EOM scheme of Refs.~\onlinecite{TserkovnikovTheorMathPhys81,TserkovnikovTheorMathPhys99}
on the Keldysh contour. This generalization paves a way 
to application of the scheme to non-equilibrium systems.
As expected, the chain of equations (\ref{EOMirrGleft})
and (\ref{EOMirrGright}) is usually infinite. However, contrary to 
the standard EOM scheme, {\it this normalized irreducible formulation 
guarantees that truncating the chain at step $n$ is equivalent 
to neglecting higher order correlations only}.\footnote{A detailed discussion and graphic representation for equilibrium Green functions EOMs can be found in Ref~\onlinecite{TserkovnikovTheorMathPhys99}.}
Note that structure of Eqs.~(\ref{EOMirrGleft}) and (\ref{EOMirrGright}) 
for $n=1$ is similar to that of the first Hubbard (HIA) 
approximation derived for the Hubbard  non-equilibrium Green functions 
(see e.g. Ref.~\onlinecite{SandalovPRB07}).

Similar to usual EOM techniques each next step takes into account higher correlations induced in the system by coupling (in our case) to baths. Note however that in the standard EOM techniques each next step brings in addition to higher order processes also multiple correlations of the lower order. In the present methodology those are excluded by the orthogonalization procedure. A detailed discussion and graphic representation for equilibrium Green functions EOMs can be found in Ref.~\onlinecite{TserkovnikovTheorMathPhys99}.

\subsection{EOM for canonical Green functions}\label{methodB}
While the form of Eqs.~(\ref{EOMirrGleft}) and (\ref{EOMirrGright}) is suggestively of
a Dyson-type, the quasiparticles described by the corresponding Green functions
are of a non-canonical nature due to spectral weight $\mathbf{P}_{A_n,A_n}(\tau)$
in the right side of the equations.\cite{ShastryPRB10} 
Also left vs. right symmetry is not obvious in the present form of the equations.
A local gauge transformation with a space-time factor is needed to achieve canonical form
of the expressions. To reveal canonical quasiparticles we choose the following scaling
\begin{equation}
 \label{defcqpG}
 \mathcal{G}^{(n-1)}_{A_n,A_n}(\tau,\tau') \equiv
 \mathbf{P}_{A_n,A_n}^{-1/2}(\tau)\,\mathbf{G}^{(n-1)}_{A_n,A_n}(\tau,\tau')\,
 \mathbf{P}_{A_n,A_n}^{-1/2}(\tau')
\end{equation}
Applying transformation (\ref{defcqpG}) to the left EOM, Eq.~(\ref{EOMirrGleft}),
leads to an EOM for the canonical Green function
\begin{align}
 \label{EOMcqpG}
&\left[i\frac{\overset{\rightarrow}{\partial}}{\partial\tau}-\mathcal{W}_n(\tau)\right]
\mathcal{G}^{(n-1)}_{A_n,A_n}(\tau,\tau') =
 \delta(\tau,\tau')\, \mathbf{I}_n
 \\ &\qquad
 + \int_c d\tau_1\,\mathcal{S}^{(n-1)}_{A_n,A_n}(\tau,\tau_1)\,
 \mathcal{G}^{(n-1)}_{A_n,A_n}(\tau_1,\tau')
 \nonumber 
\end{align}
Here
\begin{align}
 \label{defcqpS}
 &\mathcal{S}^{(n-1)}_{A_n,A_n}(\tau,\tau') =
 \mathbf{P}_{A_n,A_n}^{1/2}(\tau)\,\mathbf{\Sigma}^{(n-1)}_{A_n,A_n}(\tau,\tau')\,
 \mathbf{P}_{A_n,A_n}^{1/2}(\tau')
\nonumber \\ &\quad
 \equiv
 \mathcal{N}_{n,n+1}(\tau)\,\mathcal{G}^{(n)}_{A_{n+1},A_{n+1}}(\tau,\tau')\,
 \mathcal{N}_{n+1,n}(\tau'),
\end{align}
with
\begin{equation}
 \label{defcqpN}
 \mathcal{N}_{n,n+1}(\tau)\equiv \mathbf{P}_{A_n,A_n}^{-1/2}(\tau)\,
 \mathbf{\nu}_{n,n+1}(\tau)\,\mathbf{P}_{A_{n+1},A_{n+1}}^{1/2}(\tau)
\end{equation}
and $\mathcal{N}_{n+1,n}(\tau)=\mathcal{N}_{n,n+1}^\dagger(\tau)$,
is the canonical quasiparticle self-energy. To derive the second line in Eq.(\ref{defcqpS})
we employed Eqs.~(\ref{defS}) and (\ref{defcqpG}). 
The free evolution matrix is (see Appendix~\ref{appB} for derivation)
\begin{align}
 \label{defcqpW}
 \mathcal{W}_n(\tau) = &
 \frac{i}{2}\frac{\partial}{\partial\tau}\left[\mathbf{P}_{A_n,A_n}^{1/2}(\tau)\right]\,
 \mathbf{P}_{A_n,A_n}^{-1/2}(\tau) 
 \\ +&
 \frac{1}{2}
 \mathbf{P}_{A_n,A_n}^{-1/2}(\tau)\,\mathbf{\omega}_n(\tau)\,
 \mathbf{P}_{A_n,A_n}^{1/2}(\tau) + H.c.
 \nonumber
 \end{align}
 Note that since the free evolution matrix is Hermitian, 
 $\mathcal{W}_n(\tau)=\mathcal{W}_n^\dagger(\tau)$,
and the self-energy matrix $\mathcal{S}^{(n-1)}_{A_n,A_n}(\tau,\tau')$  is symmetric 
relative to normalization factors, Eq.(\ref{defcqpS}), a similar derivation starting 
right EOM, Eq.(\ref{EOMirrGright}), leads to the same result. 
Thus Eq.(\ref{EOMcqpG}) is of canonical Dyson form.
This is the main {\em exact} result of our consideration.

Calculations based on  the infinite chain of EOMs, Eq.(\ref{EOMcqpG}), are truncated at some finite
step $n$ by an approximation performed on self-energy 
$\mathcal{S}^{(n-1)}_{A_n,A_n}(\tau,\tau')$  of the last equation in the chain.
For example, one can completely neglect the self-energy, or assume a decoupling
procedure, which will allow expressing the self-energy in terms of Green functions
available in the chain. The resulting finite set of equations has to be solved self-consistently,
since spectral weights (normalization factors), Eq.(\ref{defP}), can be expressed in terms
of Green functions, which they also define (see Eqs.~(\ref{EOMcqpG})-(\ref{defcqpW})). 

Note that while (similar to the standard EOM methods) truncation of the chain (\ref{defcqpG}) cannot formally guarantee conserving 
character of the approximation, the methodology described here does resolve two 
important drawbacks of the usual EOM formulations: 1.~Contrary to the usually employed truncation schemes it provides  a way to make decoupling controllable in a sense that 
at each step one takes into account  systematically the greater complexity of the correlations;
2.~It resolves the symmetry-breaking problem in decoupling
chain of Green function EOMs in the sense of left vs. right EOM incompatibility. 
The latter problem  is relevant for EOM formulations for both usual\cite{BonchBruevich_Tyablikov_1962,LevyRabaniJCP13,LevyRabaniJPCM13} 
and Hubbard\cite{SandalovIJQC03,FranssonPRB05,SandalovPRB07,GalperinNitzanRatnerPRB08} Green functions, and may lead to complications (for example, non-Hermiticity of the density matrix). In the presented formulation Hermiticity and positive definiteness 
of the density matrix is guaranteed by the canonical from of Eq.(\ref{defcqpG}) 
(see Appendix~\ref{app2}).
Note also that in the examples presented in Section~\ref{numres} the approach provides results which do not violate conservation laws.
Finally, infinite chain of EOMs (\ref{defcqpG}) is exact, and can be viewed as 
a way to define proper quasiparticles for strongly interacting molecules in junctions.
Note that these quasiparticles are not are not 
molecular or Kohn-Sham orbitals (even when the latter are obtained from an (unknown)
exact pseudopotential).

\subsection{Application to transport in junctions}\label{methodC}
Till now our considerations were exact. However,
direct application of the scheme described above to open systems leads to a set
of hierarchical EOMs. Each subsequent step in the hierarchy accounts for additional
correlations caused by system-bath coupling. The resulting multi-energy correlation
functions are similar to those discussed in the literature\cite{YanJCP08}
with the distinction that in the present consideration the hierarchy is formulated 
for Green functions rather than for a density matrix. 
While the scheme is well defined, and in combination with TDDFT may be even 
practically applicable\cite{YanJCP10} (see Section~\ref{numresA} below), 
rigorous many-body formulation quickly becomes prohibitively expensive. 

A simple approximate 
practical alternative is based on substitution of the infinite chain of equations,
Eq.~(\ref{EOMcqpG}), by the first equation in the chain ($n=1$) with self-energy 
$\mathcal{S}^{(0)}_{A_1,A_1}(\tau,\tau')$ obtained from an approximate
self-consistent formulation. 
We note that the self-energy  $\mathcal{S}^{(0)}_{A_1,A_1}(\tau,\tau')$ is defined 
in terms of irreducible Green function $\mathcal{G}_{A_2,A_2}^{(1)}(\tau,\tau')$,
Eq.(\ref{defcqpS}). Then utilizing Eqs.~(\ref{defAn}), (\ref{defAnd}), (\ref{defirrG}),
and (\ref{definvop}) after lengthy but straightforward transformations we get 
(see Appendix~\ref{appC} for derivation)
\begin{align}
 \label{EOMcqpS}
& \mathcal{S}^{(0)}_{A_1,A_1}(\tau,\tau') = \mathcal{S}(\tau,\tau')
 -\mathcal{R}(\tau)\,\mathcal{G}^{(0)}_{A_1,A_1}(\tau,\tau')\,
 \mathcal{R}^{\dagger}(\tau')
\nonumber \\ &
 -\int_c d\tau_1\bigg( 
 \mathcal{R}(\tau)\,
 \mathcal{G}^{(0)}_{A_1,A_1}(\tau,\tau_1)\,
 \mathcal{S}_{A_1,A_1}^{(0)}(\tau_1,\tau')
\nonumber \\ & \qquad\quad
 +\mathcal{S}_{A_1,A_1}^{(0)}(\tau,\tau_1)\,\mathcal{G}^{(0)}_{A_1,A_1}(\tau_1,\tau')\,
 \mathcal{R}^{\dagger}(\tau')
 \bigg)
 \\ &
 -\int_c d\tau_1\int_c d\tau_2\,\mathcal{S}_{A_1,A_1}^{(0)}(\tau,\tau_1)\,
 \mathcal{G}^{(0)}_{A_1,A_1}(\tau_1,\tau_2)\,
 \mathcal{S}_{A_1,A_1}^{(0)}(\tau_2,\tau')
 \nonumber
 \end{align}
 where 
\begin{align}
\label{defcalS}
  \mathcal{S}(\tau,\tau') \equiv &
\mathbf{P}^{-1/2}_{A_1,A_1}(\tau)\,\mathbf{\nu}_{1,2}(\tau)\,
 \mathbf{G}_{a_2,a_2}(\tau,\tau')\,
 \\ &\times
 \mathbf{\nu}_{2,1}(\tau')\,\mathbf{P}_{A_1,A_1}^{-1/2}(\tau')
\nonumber \\
 \label{defR}
  \mathcal{R}(\tau) \equiv & \mathbf{P}_{A_1,A_1}^{-1/2}(\tau)\,
  \mathbf{\nu}_{1,2}(\tau)\,\mathbf{P}_{a_2,A_1}(\tau)\,
  \mathbf{P}_{A_1,A_1}^{-1/2}(\tau)
 \end{align}
 So far derivations are exact, that is the infinite chain of equation (\ref{EOMcqpG})
 can be exactly replaced by the first equation from the chain and Eq.(\ref{EOMcqpS}).
 Note that the latter system of two equations should be solved self-consistently since
 Green function $\mathcal{G}^{(0)}_{A_1,A_1}(\tau,\tau')$ is defined by self-energy
 $\mathcal{S}^{(0)}_{A_1,A_1}(\tau,\tau')$, Eq.(\ref{EOMcqpG}) with $n=1$, 
 which in turn depends on the Green function, Eq.(\ref{EOMcqpS}).
 
 However, the resulting scheme cannot be rigorous due to the unknown term
 $\mathcal{S}(\tau,\tau')$ in the right side of 
 Eq.(\ref{EOMcqpS}). This term is defined with usual Green function 
 $\mathbf{G}_{a_2,a_2}(\tau,\tau')$, Eq.(\ref{defcalS}), so that an approximation 
 representing the latter in terms of known quantities (for example, separation
 between system and bath coordinates) completes the formulation of a simple 
 practical scheme. After the approximation the self-energy 
 $\mathcal{S}^{(0)}_{A_1,A_1}(\tau,\tau')$ can be obtained from Eq.(\ref{EOMcqpS}), 
 for example within a self-consistent procedure, which starts by substitution of 
 $\mathcal{S}(\tau,\tau')$ in place of $\mathcal{S}^{(0)}_{A_1,A_1}(\tau,\tau')$ 
 in the right side of the expression. 
 
 Note that if initial vector $\hat a_1$ consists of all possible excitations in the system,
 the standard non-equilibrium Green function (correlation function of
 operators of elementary excitations) can be expressed as a combination
 of Green functions $\mathbf{G}^{(0)}_{A_1,A_1}(\tau,\tau')$. 
 The latter can be used in the standard NEGF expression for 
 current.\cite{JauhoWingreenMeirPRB94}

\section{Numerical results}\label{numres}
We apply the methodology described above to several models, and compare results 
with those obtained within other techniques.

\subsection{Non-interacting multi-level system}\label{numresA}
As a starting point we consider a non-interacting system of $M$ levels (orbitals)
coupled to contacts. The Hamiltonian of the model is
\begin{align}
 \label{HA}
 \hat H =& \sum_{m_1,m_2=1}^M H^{(M)}_{m_1m_2}\hat d_{m_1}^\dagger\hat d_{m_2}
 +\sum_k \varepsilon_k\hat c_k^\dagger\hat c_k
 \\ & \qquad\qquad +
 \sum_{k,m}\left(V_{mk}\hat d_m^\dagger\hat c_k + H.c.\right)
 \nonumber
\end{align}
Here first and second terms represent free electrons in the system and baths, respectively.
Third term is the bi-linear coupling between them. 

Since we are interested in the properties of 
the system, it is natural to choose molecular quasiparticle excitations as the initial vector operator
\begin{equation}
\hat a_1 \equiv \hat A_1 = \left(\left\{\hat d_m\right\}\right)
\end{equation}
With this choice we get
\begin{equation}
 \hat A_2 = \left(\left\{\hat c_k\right\}\right) \qquad
 \hat A_3 = 0
\end{equation}
and the chain of equations closes exactly at the second step. 
Moreover, due to the standard commutation relations $\mathbf{P}_{A_1,A_1}=\mathbf{I}_M$
and $\mathbf{P}_{a_2,A_1}=0$, so that the canonical quasi-particle Green functions,
$\mathcal{G}^{(0)}_{A_1,A_1}(\tau,\tau')$, and self-energies,
$\mathcal{S}^{(0)}_{A_1,A_1}(\tau,\tau')$, coincide with the standard non-equilibrium
Green function formulations, and Eq.(\ref{EOMcqpG}) (with $n=1$) is the standard Dyson 
equation.

\begin{figure}[htbp]
\centering\includegraphics[width=\linewidth]{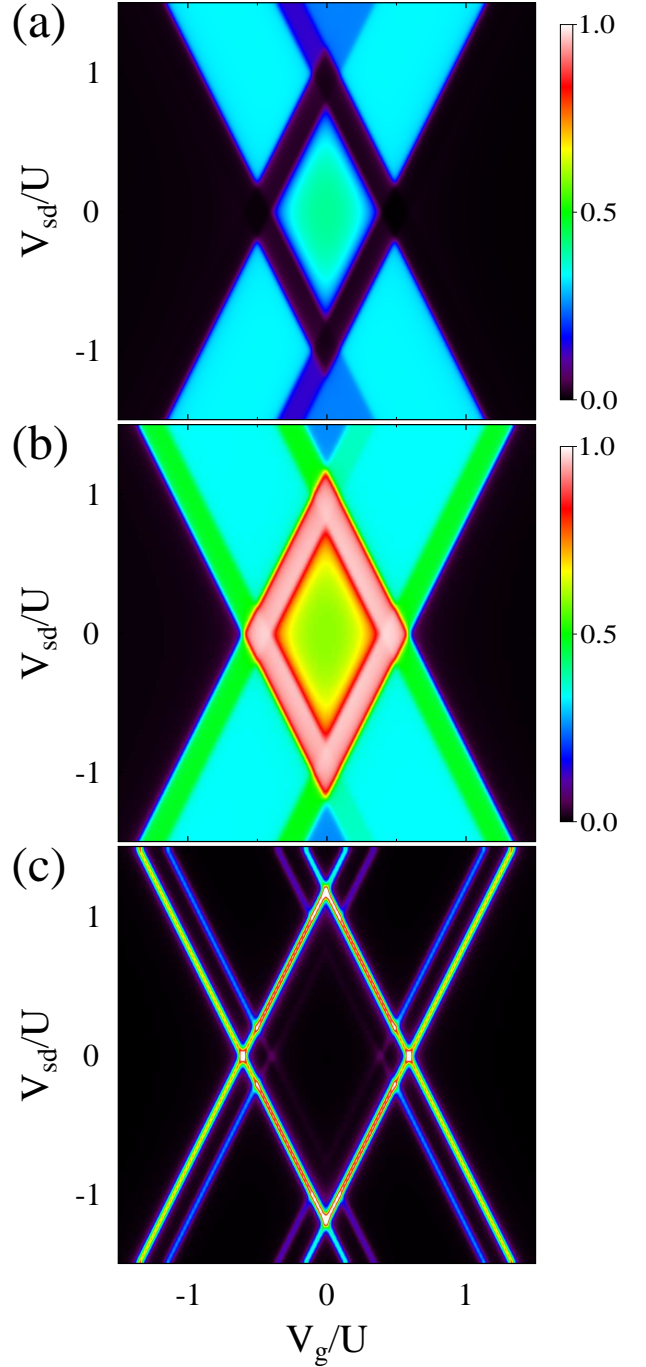}
\caption{\label{fig1}
(Color online) QD in Coulomb blockade regime. Shown are maps of
(a) probability of state $\lvert\uparrow\rangle$, $P_\uparrow$; 
(b) probability of state $\lvert\downarrow\rangle$, $P_\downarrow$; 
and (c) conductance, $dI/dV_{sd}$. See text for parameters.
}
\end{figure}

\subsection{Coulomb blockade in QD}\label{numresB}
Next we turn to a quantum dot in a junction with paramagnetic leads.
The model Hamiltonian is
\begin{align}
\label{HB}
 \hat H =& \sum_{\sigma}\varepsilon_\sigma\hat n_\sigma
 + U\hat n_\uparrow\hat n_{\downarrow}
 +\sum_{k,\sigma} \varepsilon_{k}\hat c_{k\sigma}^\dagger\hat c_{k\sigma}
 \\
 +& \sum_{k,\sigma}\left(V_k\hat d_\sigma^\dagger\hat c_{k\sigma} + H.c.\right)
 \nonumber
\end{align}
where $\varepsilon_\sigma=\varepsilon-\sigma g\mu_B B_{dc}/2$
($g$  is the Lande factor, $\mu_B$ is the Bohr magneton, $B_{dc}$ is the amplitude
of {\em dc} magnetic field), $\hat n_\sigma=\hat d_\sigma^\dagger\hat d_\sigma$, 
and U is Coulomb  repulsion term.

Elementary system excitations (as in Section~\ref{numresA}) are not the best
choice for the initial vector operator. Such a starting point will lead to a Hartree-type
treatment at the first step of the EOM chain. For strong $U$ a more suitable choice
is a non-equilibrium atomic limit, where state resolved system excitations
(Hubbard operators) form $\hat a_1$
\begin{equation}
 \hat a_1\equiv\hat A_1 = \left((1-\hat n_\downarrow)\hat d_\uparrow,\,
 (1-\hat n_\uparrow)\hat d_\downarrow,\,
 \hat n_\downarrow\hat d_\uparrow,\,\hat n_\uparrow\hat d_\downarrow\right)^T
\end{equation}
where $(\ldots)^T$ is transpose operation. Its spectral weight and free evolution 
matrices are
\begin{align}
 &\mathbf{P}_{A_1,A_1} = \mbox{diag}\left[
 1-\langle\hat n_\downarrow\rangle,\, 1-\langle \hat n_\uparrow\rangle,
 \langle\hat n_\downarrow\rangle,\, \langle \hat n_\uparrow\rangle
 \right]
 \\
& \mathbf{\varepsilon}_1 = \mbox{diag}\left[
 \varepsilon_\uparrow,\,\varepsilon_\downarrow,\,
 \varepsilon_\uparrow+U,\,\varepsilon_\downarrow+U
 \right]
\end{align}
Here we want to restrict consideration to
the Coulomb blockade regime, i.e. all the bath-assisted spin-spin correlations can be neglected
(these correlations are important for the Kondo effect, as discussed below in 
Section~\ref{numresC}). This leads to
\begin{align}
& \hat A_2 = \left((1-\hat n_\downarrow)\hat c_{k\uparrow},\,
 (1-\hat n_\uparrow)\hat c_{k\downarrow},\,
 \hat n_\downarrow\hat c_{k\uparrow},\,\hat n_\uparrow\hat c_{k\downarrow}\right)^T
 \\
&\mathbf{P}_{A_2,A_2} = \mbox{diag}\left[
 1-\langle\hat n_\downarrow\rangle,\, 1-\langle \hat n_\uparrow\rangle,
 \langle\hat n_\downarrow\rangle,\, \langle \hat n_\uparrow\rangle
 \right]
 \\
& \mathbf{\varepsilon}_2 = \mbox{diag}\left[
 \varepsilon_k,\,\varepsilon_k,\,\varepsilon_k+U,\,\varepsilon_k+U
 \right]
\end{align}  
and $\mathcal{W}_1\equiv\mathbf{\varepsilon}_1$.

Truncating the EOM chain (\ref{EOMcqpG}) at the second step by neglecting 
self-energy $\mathcal{S}^{(1)}_{A_2,A_2}$ we get a Dyson equation for
canonical Green function $\mathcal{G}^{(0)}_{A_1,A_1}(\tau,\tau')$
with self-energy
\begin{align}
\label{CBS0}
\mathcal{S}^{(0)}_{A_1,A_1}(\tau,\tau') =&
\mbox{diag}\bigg[
 \Sigma_\uparrow(\tau,\tau'),\,\Sigma_\downarrow(\tau,\tau'),\,
 \\ &\qquad\quad
  \Sigma_\uparrow(\tau,\tau'),\,\Sigma_\downarrow(\tau,\tau')
 \bigg]
 \nonumber
\end{align}
where $\Sigma_\sigma(\tau,\tau')\equiv\sum_k |V_k|^2 g_{k\sigma}(\tau,\tau')$,
and $g_{k\sigma}(\tau,\tau')\equiv -i\langle T_c\,\hat c_{k\sigma}(\tau)\,\hat c_{k\sigma}^\dagger(\tau')\rangle$ is the Green function of free electrons in the contacts.

Utilizing relation $\hat d_\sigma=(1-\hat n_{\bar\sigma})\hat d_\sigma+\hat n_{\bar\sigma}\hat d_\sigma$ (here $\bar\sigma$ is spin projection opposite to $\sigma$), 
one can express the standard non-equilibrium Green function
$G_{\sigma}(\tau,\tau')\equiv-i\langle T_c\,\hat d_\sigma(\tau)\,\hat d_\sigma^\dagger(\tau')\rangle$ as a combination of Hubbard non-equilibrium Green functions $\mathbf{G}^{(0)}_{A_1,A_1}(\tau,\tau')$. The resulting expression is
\begin{equation}
\label{NEGF_G}
 G_\sigma(\tau,\tau') = \left[1-\langle\hat n_{\bar\sigma}\rangle\right] G(\tau,\tau')
 + \langle\hat n_{\bar\sigma}\rangle G_U(\tau,\tau')
\end{equation}
where $G(\tau,\tau')$ and $G_U(\tau,\tau')$ are non-equilibrium Green functions defined by
the following Dyson equations 
\begin{align}
 &\left[i\frac{\partial}{\partial\tau}-\varepsilon_\sigma\right] G(\tau,\tau') = \delta(\tau,\tau')
 \\ &\qquad\qquad
 + \int_c d\tau_1\,\Sigma_{\sigma}(\tau,\tau_1)\,G(\tau_1,\tau')
\nonumber \\
 &\left[i\frac{\partial}{\partial\tau}-\varepsilon_\sigma-U\right] G_U(\tau,\tau') = \delta(\tau,\tau')
 \\ &\qquad\qquad
 + \int_c d\tau_1\,\Sigma_{\sigma}(\tau,\tau_1)\,G_U(\tau_1,\tau')
 \nonumber
\end{align}
We note that Eq.(\ref{NEGF_G}) coincides with Eq.(21) of 
Ref.~\onlinecite{GalperinNitzanRatnerPRB07}, where it was obtain from the standard EOM approach. However, while a special symmetrization procedure
had to be implemented in the latter case (see Eq.(A23) of Ref.~\onlinecite{GalperinNitzanRatnerPRB07}) here the same expression
is obtained in the symmetrized form automatically. This is an example of 
the built-in symmetrization property of the formulation, which otherwise 
should be obtained within a step-by-step approach.\cite{LevyRabaniJCP13,LevyRabaniJPCM13}
Note also that in the Coulomb blockade regime the EOM for the canonical Green function 
$\mathcal{G}^{(0)}_{A_1,A_1}(\tau,\tau')$ can be solved directly (without a self-consistent
procedure). This is reminiscent of the similar observation in Ref.~\onlinecite{GalperinNitzanRatnerPRB07} (see discussion below Eq.(43) there).

The spectral weight $\mathbf{P}_{A_1,A_1}$ is calculated from the lesser projections
of the canonical Green function $\mathcal{G}^{(0)}_{A_1,A_1}$ using
\begin{align}
 n_\uparrow=& \frac{(1-I_2)\,I_1+I_2\,I_3}{1-(I_1-I_3)(I_2-I_4)}
 \\
 n_\downarrow=& \frac{(1-I_1)\,I_2+I_1\,I_4}{1-(I_1-I_3)(I_2-I_4)}
\end{align}
where $I_m \equiv -i \mathcal{G}^{(0)\, <}_{A_1^{(m)},A_1^{(m)}}(t,t)$
($A_1^{(m)}$ indicates $m^{th}$ operator of the vector $\hat A_1$).

In addition to the level populations, probabilities of many-body states
of the quantum dot, $\lvert 0\rangle$, $\lvert\sigma\rangle$, and $\lvert 2\rangle$,
can be evaluated from 
\begin{align}
 &P_\uparrow=-i\mathbf{G}^{(0)\, <}_{A_1^{(1)},A_1^{(1)}}(t,t)\quad
 P_\downarrow=-i\mathbf{G}^{(0)\, <}_{A_1^{(2)},A_1^{(2)}}(t,t)
 \\ &
 P_2=-i\mathbf{G}^{(0)\, <}_{A_1^{(3)},A_1^{(3)}}(t,t)
       =-i\mathbf{G}^{(0)\, <}_{A_1^{(4)},A_1^{(4)}}(t,t)
 \nonumber
\end{align}
and $P_0=1-P_\uparrow-P_\downarrow-P_2$.

Figure~\ref{fig1} shows results of calculations which employ EOM for the canonical 
Green function $\mathcal{G}^{(0)}_{A_1,A_1}(\tau,\tau')$. 
Parameters are (in units of $U$): temperature $T=10^{-3}$, 
quantum dot levels $\varepsilon_\uparrow=-0.4$ and
$\varepsilon_\downarrow=-0.6$, 
electronic escape rate to contact $K$ ($K=L,R$)
$\Gamma_K\equiv 2\pi\sum_k |V_k|^2\delta(E-\varepsilon_k)=0.01$
(we assume the wide band limit, where the escape rate does not depend on energy $E$).
The Fermi energy is taken at origin $E_F=0$, and bias is applied symmetrically,
$\mu_L=E_F+V_{sd}/2$ and $\mu_R=E_F-V_{sd}/2$. Gate potential shifts level
positions $\varepsilon_\sigma\to\varepsilon_\sigma+V_g$. Calculations are performed
on an energy grid spanning the region from $-3$ to $3$ with step $10^{-4}$.
Probabilities of states $\lvert\uparrow\rangle$ and $\lvert\downarrow\rangle$
(see Figs.~\ref{fig1}a and b)
are seen to be equal in the blockaded and transition regions, but differ considerably 
at the borders of the two, where the probability of the state $\lvert\downarrow\rangle$
is much higher due to the lower position of its energy level.
Fig.~\ref{fig1}c shows a conductance map with spin sidebands.
The result  is in agreement with experimental observation.\cite{Park2002}

\begin{figure}[t]
\centering\includegraphics[width=\linewidth]{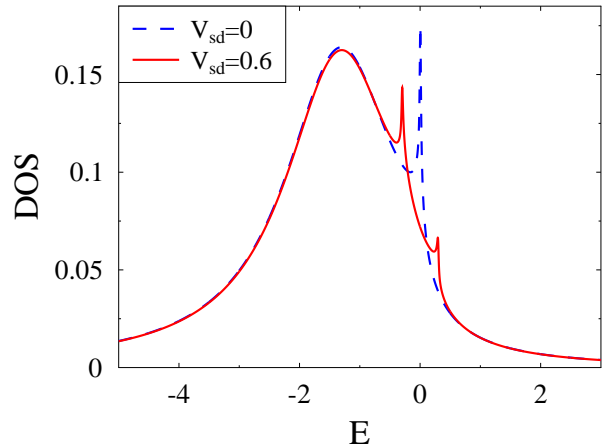}
\caption{\label{fig2}
(Color online) Kondo in QD. Density of states, $-\mbox{Im}\, G^r_\sigma(E)/\pi$,
in equilibrium (dashed line, blue) and biased $V_{sd}=0.6$ (solid line, red)
junction. See text for parameters.
}
\end{figure}
\subsection{Kondo in QD}\label{numresC}
We continue consideration of the QD model, Eq.(\ref{HB}).
Note that any nonequilibrium atomic limit (by its construction taking all interactions 
in the system into account exactly, but treating system-bath coupling perturbatively) 
is not a proper starting point to describe effects of the Kondo type,
where entanglement between the system and bath has to be taken into account 
in a non-perturbative manner. Thus it is not reasonable to expect that the results
presented below reproduce Kondo features exactly. So we compare our treatment
to standard EOM approaches available in the literature.\cite{GalperinNitzanRatnerPRB07,MeirWingreenLeePRL93,ZhangXueXiePRL03,HattoriPRB08,FranssonZhuPRB08}

To obtain Kondo features, one has to truncate the EOM chain at least on the third step.\cite{GalperinNitzanRatnerPRB07,MeirWingreenLeePRL93}
Correspondingly, in the present approach one has to work with three-dimensional 
energy grid. To avoid heavy computations we make the following simplifications:
1.~Kondo is observed for $U\to\infty$, thus all Hubbard operators corresponding to
double occupation of QD can be neglected; 2.~The effect describes coherence between 
system and bath, thus excitations in both should be accounted for from the start;
3.~Correlations responsible for appearance of a Kondo peak include bath-induced 
spin-spin correlations in the system, thus to simplify
the consideration we include excitations of this type into the initial vector 
operator $\hat a_1$; 4.~For simplicity we neglect system-induced spin-spin correlations
in the bath.

With these approximations in mind an initial choice is 
\begin{equation}
\label{KA1}
 \hat A_1 = \bigg((1-\hat n_{\bar\sigma})\hat d_\sigma,\,
 (1-\hat n_{\bar\sigma})\hat c_{k\sigma},\,
 \hat d_{\bar\sigma}^\dagger\hat c_{k\bar\sigma}\hat d_\sigma
 \bigg)^T
\end{equation}
where $\sigma=\uparrow,\downarrow$, 
$\hat n_\sigma\equiv\hat d_\sigma^\dagger\hat d_\sigma$, and
$\hat n_{k\sigma}\equiv\hat c_{k\sigma}^\dagger\hat c_{k\sigma}$.
Corresponding spectral weight and free evolution matrices have block spin structure
\begin{align}
&\mathbf{P}_{A_1,A_1}^{(\sigma)}=
\\ &
\begin{bmatrix}
\langle 1-\hat n_{\bar\sigma}\rangle & 0 & 
\langle(1-\hat n_\sigma)\hat c_{k\bar\sigma}^\dagger\hat d_{\bar\sigma}\rangle 
\\
0 & \langle 1-\hat n_{\bar\sigma}\rangle &
\langle\hat c_{k\bar\sigma}^\dagger\hat c_{k\sigma}\,\hat d_\sigma^\dagger\hat d_{\bar\sigma}\rangle
\\
\langle(1-\hat n_\sigma)\hat d^\dagger_{\bar\sigma}\hat c_{k\bar\sigma}\rangle &
\langle\hat c_{k\sigma}^\dagger\hat c_{k\bar\sigma}\,\hat d_{\bar\sigma}^\dagger\hat d_\sigma\rangle &
\langle \hat n_{\bar\sigma}+(\hat n_\sigma-\hat n_{\bar\sigma})\hat n_{k\bar\sigma}\rangle  
\end{bmatrix}
\nonumber \\ &
\mathbf{\varepsilon}_1^{(\sigma)}=
\begin{bmatrix}
\varepsilon_\sigma & V_{k\sigma} & -V_{k\bar\sigma} 
\\
\overset{*}{V}_{k\sigma} & \varepsilon_{k\sigma} & 0
\\
0 & 0 & \varepsilon_{k\bar\sigma}-\varepsilon_{\bar\sigma}+\varepsilon_\sigma
\end{bmatrix}
\end{align}
Neglecting system-induced spin correlations in the bath we get
\begin{align}
&\hat a_2 = \bigg(\hat n_{k\bar\sigma}\hat d_\sigma\bigg)^T
\\ &
\mathbf{P}_{a_2,A_1}^{(\sigma)} =
\begin{bmatrix}
\langle(1-\hat n_{\bar\sigma})\hat n_{k\bar\sigma}\rangle & 0 &
\langle(1-\hat n_\sigma)\hat c^\dagger_{k\bar\sigma}\hat d_{\bar\sigma}\rangle
\end{bmatrix}
\end{align}

To allow for analytic derivation we make an additional approximation, 
decoupling system and bath degrees of freedom in $\mathbf{P}_{A_1,A_1}$
and $\mathbf{P}_{a_2,A_1}$. Note that this type of approximations should be 
done with care. In particular, derivation of the symmetrized version for free evolution
presented in Section~\ref{methodB} is not possible after the approximation,
and free evolution term in Eq.\ (\ref{EOMcqpG}), should be taken in the form 
presented in Eq.(\ref{Bstart}). Fortunately, in this particular case the resulting EOM
for the system part of the canonical Green function 
(canonical correlation function of the first operator in  the vector $\hat A_1$, Eq(\ref{KA1})) 
is the same if derived from left or right EOM. Explicit expression for the EOM is
\begin{align}
\label{KcqpG}
&\left(i\frac{\partial}{\partial\tau}-\varepsilon_\sigma\right)
\mathcal{G}^{(0)}_{A_1^{(1)},A_1^{(1)}}(\tau,\tau') = \delta(\tau,\tau')
\\ &
+\int_c d\tau_1\left(\Sigma_\sigma(\tau,\tau_1)+\Sigma_\sigma^{(K)}(\tau,\tau_1)\right)
\mathcal{G}^{(0)}_{A_1^{(1)},A_1^{(1)}}(\tau_1,\tau') 
\nonumber
\end{align}
where self-energy $\Sigma_\sigma(\tau,\tau')$ is defined below Eq.(\ref{CBS0}) and
\begin{equation}
\label{KSK}
 \Sigma_\sigma^{(K)}(\tau,\tau') \equiv \sum_k \lvert V_k\rvert^2 
 \langle\hat n_{k\bar\sigma}\rangle g_{k\bar\sigma}^{(K)}(\tau,\tau')
\end{equation}
with 
\begin{equation}
\left(i\frac{\partial}{\partial\tau}-\varepsilon_{k\bar\sigma}+\varepsilon_{\bar\sigma}-\varepsilon_\sigma\right) g_{k\bar\sigma}^{(K)}(\tau,\tau') = \delta(\tau,\tau')
\end{equation}
defining the contact electron Green function.
The latter self-energy is responsible for appearance of Kondo features.
We note that the corresponding standard non-equilibrium Green function,
$G_\sigma(\tau,\tau')\equiv (1-\langle\hat n_{\bar\sigma}\rangle)\mathcal{G}^{(0)}_{A_1^{(1)},A_1^{(1)}}(\tau,\tau')$ coincides with the result obtained in our previous
publication (see Eq.(53) in Ref.~\onlinecite{GalperinNitzanRatnerPRB07}).
Note also that element $\langle\hat n_{k\bar\sigma}\rangle$ in Eq.(\ref{KSK}),
central for appearance of Kondo features, is obtained as a result of projection
(orthogonalization), Eqs.~(\ref{defAn}) and (\ref{defAnd}).  

Figure~\ref{fig2} shows results of calculations that employ EOM for the canonical 
Green function $\mathcal{G}^{(0)}_{A_1^{(1)},A_1^{(1)}}(\tau,\tau')$, Eq.(\ref{KcqpG}). 
Parameters are (in units of $\Gamma$):
$T=0.005$, $\varepsilon_\sigma=-2$, and $\Gamma_L=\Gamma_R=0.5$.
Fermi energy is taken at origin, $E_F=0$, and bias is applied symmetrically,
$\mu_L=E_F+V_{sd}/2$ and $\mu_R=E_F-V_{sd}/2$.
Calculations are performed on an energy grid spanning the region from $-5$ to $3$ 
with step $10^{-4}$. The density of states (same for both spins) in equilibrium 
demonstrates a Kondo feature at $E_F$. Under bias the feature splits into two peaks 
centered around each of the chemical potentials. This result was first reported in
Ref.~\onlinecite{MeirWingreenLeePRL93}.

\begin{figure}[htbp]
\centering\includegraphics[width=\linewidth]{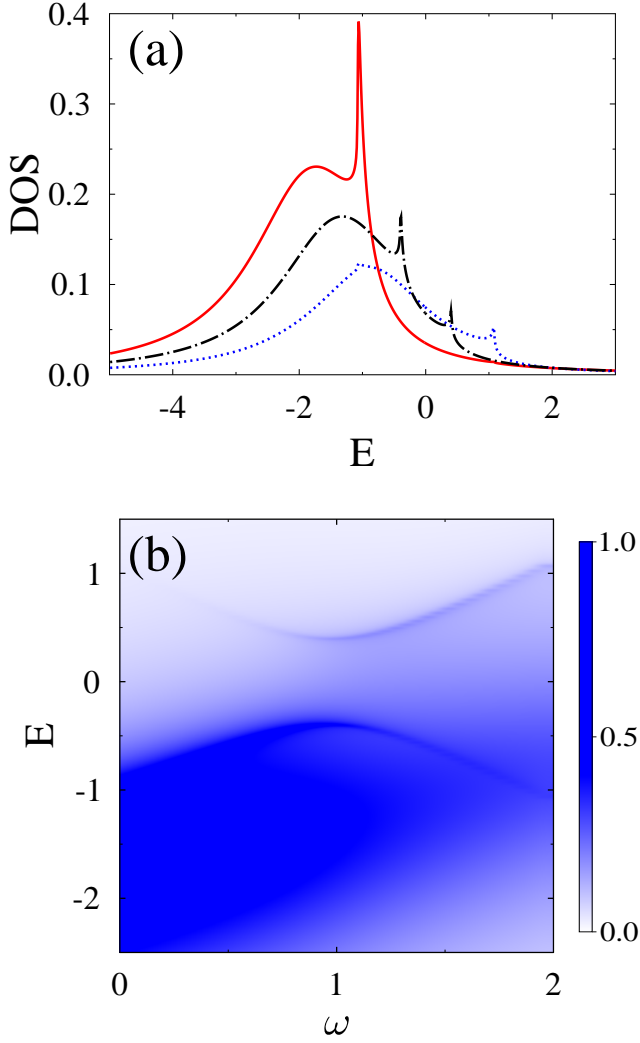}
\caption{\label{fig3}
(Color online) Kondo in QD in the presence of {\em ac} magnetic field with frequency $\omega$. 
(a) Density of states, $\rho_\sigma(E,\omega)\equiv-\mbox{Im}\left[\mathbf{U}\,\mathbf{G}^{(0)\,r}_{A_1^{(1)},A_1^{(1)}}(E)\,\mathbf{U}^\dagger\right]_{\sigma\sigma}/\pi$
for $\omega=0, \sigma=\uparrow$ (solid line, red), 
$\omega=0, \sigma=\downarrow$ (dotted line, blue), and
$\omega=1$ (dash-dotted line, black). 
(b) Map of $\rho_\uparrow(E,\omega)$. See text for parameters.
}
\end{figure}

In the presence of an {\em ac} magnetic field Hamiltonian (\ref{HB}) acquires an additional
time-dependent term
\begin{equation} 
-g\mu_B B_{ac}\left(\hat d^\dagger_\uparrow\hat d_\downarrow e^{i\omega t}
+\hat d_\downarrow\hat d_\uparrow e^{-i\omega t}\right)
\end{equation}
Here $B_{ac}$ is the amplitude and $\omega$ is the frequency of the {\em ac} magnetic field. 
Transformation to the rotational frame of the field 
\begin{equation}
\begin{aligned}
&\hat{\bar H}=i\left(\frac{\partial}{\partial t}e^{\hat S(t)}\right)e^{-\hat S(t)}
 + e^{\hat S(t)}\hat He^{-\hat S(t)} \\
&\hat S(t)=-\frac{i}{2}\omega t
 \sum_\sigma \sigma\left(\hat n_\sigma +\sum_k\hat n_{k\sigma}\right)
\end{aligned} 
\end{equation} 
allows us to formulate an effective time-independent problem.  
Finally, transformation into molecular eigenbasis leads to
\begin{align}
\label{HC}
&\hat H_{eig} = \sum_\sigma E_\sigma \hat c_\sigma^\dagger\hat c_\sigma
 + U\hat n^c_\uparrow\hat n^c_\downarrow
 + \sum_{k\sigma}\bar\varepsilon_{k\sigma}\hat c^\dagger_{k\sigma}\hat c_{k\sigma}
 \\
& + \sum_k \left[\left(\hat c_\uparrow^\dagger,\hat c^\dagger_\downarrow\right)
\mathbf{U}^\dagger V_k\binom{\hat c_{k\uparrow}}{\hat c_{k\downarrow}}
+\left(\hat c_{k\uparrow}^\dagger,\hat c_{k\downarrow}^\dagger\right)
V_k^{*} \mathbf{U}\binom{\hat c_\uparrow}{\hat c_\downarrow}\right]
\nonumber
\end{align}
where $\hat n^c_\sigma=\hat c_\sigma^\dagger\hat c_\sigma$, 
$\bar\varepsilon_{k\sigma}=\varepsilon_k+\sigma\omega/2$, and
\begin{align}
& \mathbf{U} = \begin{bmatrix}
\cos\theta & -\sin\theta \\
\sin\theta & \cos\theta
\end{bmatrix};
\qquad
\tan 2\theta = \frac{2\lvert g\mu_B B_{ac}\rvert}{\omega-g\mu_B B_{dc}}
\\
& E_\sigma = \varepsilon+\frac{\sigma}{2}\sqrt{(\omega-g\mu_B B_{dc})^2+4(g\mu_B B_{ac})^2}
\end{align}

Treatment within the projection operator NEGF EOM (PO-NEGF EOM) 
approach is similar to the one discussed above. Initial vector operator is
\begin{equation}
\begin{aligned}
\hat A_1 =& \bigg((1-\hat n^c_{\bar\sigma})\hat c_\sigma,\,
 (1-\hat n^c_{\bar\sigma})\hat c_{k\sigma},\,
 \hat c_{\bar\sigma}^\dagger\hat c_{k\bar\sigma}\hat c_\sigma,
 \\ &
  (1-\hat n^c_{\sigma})\hat c_{k\sigma},\,
  \hat c_{\bar\sigma}^\dagger\hat c_{k\sigma}\hat c_\sigma
 \bigg)^T
 \end{aligned}
\end{equation}
which yields in the second step
\begin{equation}
\hat a_2 = \bigg(\hat n_{k\bar\sigma}\hat c_\sigma,\,\hat n_{k\sigma}\hat c_\sigma\bigg)^T
\end{equation}
The set of approximations outlined above Eq.(\ref{KA1}) eventually leads to
\begin{align}
&\left(i\frac{\partial}{\partial\tau}\mathbf{I}-\mathbf{E}\right) 
\mathbf{G}^{(0)}_{A_1^{(1)},A_1^{(1)}}(\tau,\tau') = \delta(\tau,\tau')\mathbf{P} 
\\ & +\int_c d\tau_1
\left(\mathbf{\Sigma}^{(K0)}(\tau,\tau_1)
+\mathbf{\Sigma}^{(K1)}(\tau,\tau_1)
+\mathbf{\Sigma}^{(K2)}(\tau,\tau_1)\right)
\nonumber\\ &\qquad\qquad\times
\mathbf{G}^{(0)}_{A_1^{(1)},A_1^{(1)}}(\tau_1,\tau') 
\nonumber
\end{align}
where free evolution, Green function and self-energies are $2\times 2$ matrices in 
the spin space, $\mathbf{I}$ is the unit matrix,
$\mathbf{E}=\mbox{diag}[E_\uparrow,E_\downarrow]$, and
$\mathbf{\Sigma}^{(K0)}(\tau,\tau')=
\mathbf{U}^\dagger\mathbf{\Sigma}(\tau,\tau')\mathbf{U}$.
$\mathbf{\Sigma}(\tau,\tau')$, $\mathbf{\Sigma}^{(K1)}(\tau,\tau')$,
and $\mathbf{\Sigma}^{(K2)}(\tau,\tau')$ are the diagonal in spin space matrices
defined as
\begin{align}
\Sigma_\sigma(\tau,\tau') \equiv& \sum_k \lvert V_k\rvert^2 \bar g_{k\sigma}(\tau,\tau')
\\
\Sigma_\sigma^{(K1)}(\tau,\tau') \equiv& \cos^2\theta\sum_k \lvert V_k\rvert^2
\langle\hat n_{k\bar\sigma}\rangle\bar g^{(1)}_{k\bar\sigma}(\tau,\tau')
\\
\Sigma_\sigma^{(K2)}(\tau,\tau') \equiv& \sin^2\theta\sum_k \lvert V_k\rvert^2
\langle\hat n_{k\sigma}\rangle\bar g^{(2)}_{k\sigma}(\tau,\tau')
\end{align}
with
\begin{align}
&\left(i\frac{\partial}{\partial\tau}-\bar\varepsilon_{k\sigma}\right) \bar g_{k\sigma}(\tau,\tau') 
= \delta(\tau,\tau')
\\
&\left(i\frac{\partial}{\partial\tau}-\bar\varepsilon_{k\sigma}+\varepsilon_{\sigma}
-\varepsilon_{\bar\sigma}\right) \bar g_{k\sigma}^{(1)}(\tau,\tau') = \delta(\tau,\tau')
\\
&\left(i\frac{\partial}{\partial\tau}-\bar\varepsilon_{k\sigma}+\varepsilon_{\bar\sigma}
-\varepsilon_\sigma\right) \bar g_{k\sigma}^{(2)}(\tau,\tau') = \delta(\tau,\tau')
\end{align}
$\mathbf{P}$ is spectral weight
\begin{equation}
\mathbf{P} =\begin{bmatrix}
\langle 1-\hat n^c_\downarrow\rangle & 
\langle \hat c_\downarrow^\dagger\hat c_\uparrow\rangle \\
\langle \hat c_\uparrow^\dagger\hat c_\downarrow\rangle &
\langle 1-\hat n^c_\uparrow\rangle
\end{bmatrix}
\end{equation}
Neglecting the off-diagonal elements in the spectral weight $\mathbf{P}$ 
we recover the approximation discussed in Refs.~\onlinecite{ZhangXueXiePRL03,HattoriPRB08,FranssonZhuPRB08}.
Note that similar to consideration in Section~\ref{numresA}
and contrary to discussion in Ref.~\onlinecite{ZhangXueXiePRL03}
no self-consistent calculation is necessary within the approximation.

Figure~\ref{fig3} shows density of states, 
$\rho_\sigma(E,\omega)\equiv-\mbox{Im}\left[\mathbf{U}\,\mathbf{G}^{(0)\,r}_{A_1^{(1)},A_1^{(1)}}(E)\,\mathbf{U}^\dagger\right]_{\sigma\sigma}/\pi$,
for the Kondo feature in the presence of an {\em ac} magnetic field.
Parameters of the calculation are (in units of $\Gamma$):
$\varepsilon_\uparrow=-2.5$, $\varepsilon_\downarrow=-1.5$,
$g\mu_B B_{ac}=0.2$, and $V_{sd}=0$. Other parameters are as in Fig.~\ref{fig2}.
In this case densities for spin up and down are in general different (see solid and 
dotted lines in Fig.~\ref{fig3}a). They coincide at $\omega=1$ (dash-dotted line in
Fig.~\ref{fig3}a), where the frequency of the field is in resonance with molecular excitation,
$\omega=\lvert\varepsilon_\uparrow-\varepsilon_\downarrow\rvert$.
Fig.~\ref{fig3}b illustrates an avoided crossing demonstrated by Kondo peaks of the two
densities when the frequency of the {\em ac} magnetic field approaches molecular 
resonance.

\subsection{Two-level system}\label{numresD}
As a last example we consider a particular case of Hamiltonian (\ref{HA}) with $M=2$.
However instead of using quasiparticle excitation operators, $\hat d_m$,
we perform the analysis employing (in the spirit of the non-equilibrium atomic limit) 
Hubbard operators. The initial vector-operator is
\begin{equation}
\label{TLSA1}
\hat A_1 = \bigg(\hat X_{0a},\hat X_{0b},\hat X_{b2},\hat X_{a2}\bigg)^T
\end{equation} 
where $\hat X_{S_1S_2}\equiv\lvert S_1\rangle\,\langle S_2\rvert$ are 
projection (Hubbard) operators, and $\lvert 0\rangle$, $\lvert a\rangle$,
$\lvert b\rangle$, $\lvert 2\rangle$ are many-body states in the eigen-basis
of molecular Hamiltonian $\mathbf{H}^{(M)}$. The choice (\ref{TLSA1})
leads to
\begin{align}
 & \mathbf{P}_{A_1,A_1} =
 \\ &
 \begin{bmatrix}
 \langle\hat X_{00}+\hat X_{aa}\rangle & \langle\hat X_{ba}\rangle & 0 & 0
 \\
 \langle\hat X_{ab}\rangle & \langle\hat X_{00}+\hat X_{bb}\rangle & 0 & 0
 \\
 0 & 0 & \langle\hat X_{bb}+\hat X_{22}\rangle & \langle\hat X_{ba}\rangle
 \\
 0 & 0 & \langle\hat X_{ab}\rangle & \langle\hat X_{aa}+\hat X_{22}\rangle
 \end{bmatrix}
 \nonumber \\ &
 \mathbf{\varepsilon}_1 = \mbox{diag}\bigg[E_a,E_b,E_a,E_b\bigg]
\end{align}
We note that the First Hubbard Approximation (HIA) partly misses renormalization of the free evolution matrix
$\mathbf{\omega}_1$, Eq.(\ref{defomega}), and completely neglects the self-energy
$\mathbf{\Sigma}^{(0)}_{A_1,A_1}$, Eq.(\ref{defS}).

As discussed in Section~\ref{methodC}, direct application of the methodology
to transport problems in general is prohibitively expensive due to necessity
to work with multi-energy correlation functions. Here we implement an
approximation, where canonical self-energy is provided by solving
self-consistently Eq.(\ref{EOMcqpS}), and decoupling between 
system and bath variables is assumed in the correlation function $\mathcal{S}(\tau,\tau')$.

\begin{figure}[htbp]
\centering\includegraphics[width=\linewidth]{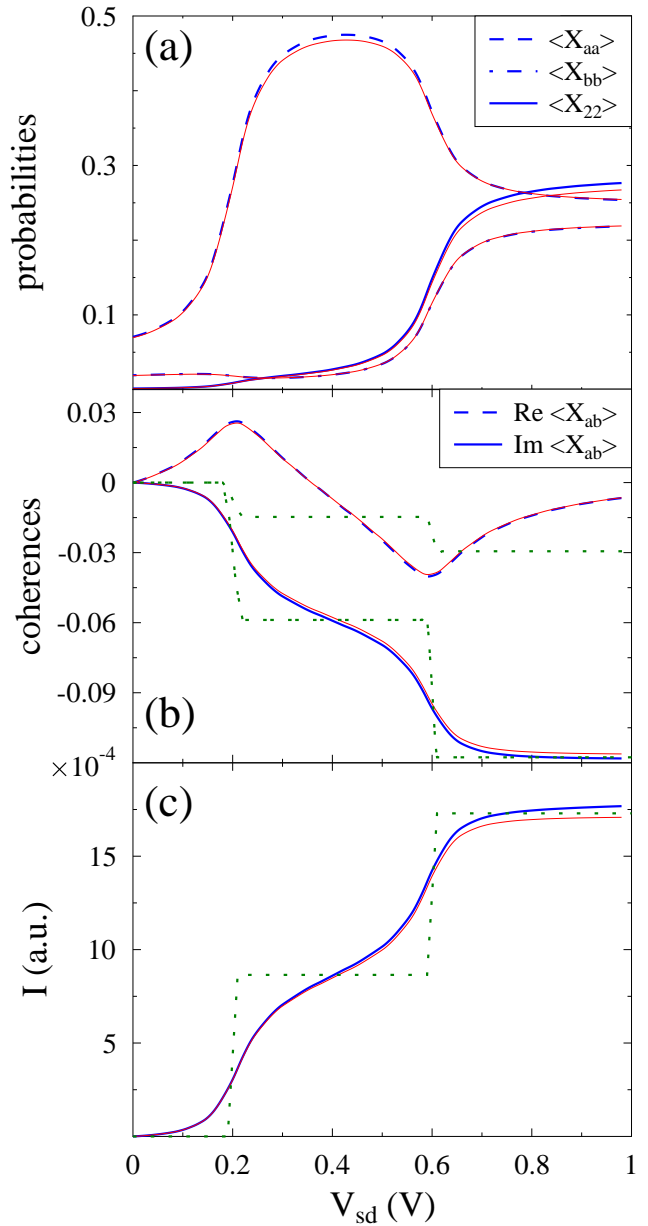}
\caption{\label{fig4}
(Color online) Two-level system (TLS). 
Simulations are performed within the NEGF (solid line; red)
and projection operator scheme of Section~\ref{methodC} 
(solid, dashed, and dash-dotted lines; blue). Shown are (a) probabilities and (b) coherences
of the many-body eigenstates of the TLS, and (c) current vs. applied bias $V_{sd}$.
Panels (b) and (c) show also results of the Redfield QME simulation 
(dotted and double-dotted lines, green). See text for parameters.
}
\end{figure}

This separation leads to the appearance of the canonical system Green functions
$\mathcal{G}^{(0)}_{B_1,B_1}(\tau,\tau')$ generated by vector-operator
\begin{equation}
\label{TLSB1}
\hat B_1 = \bigg(\hat X_{00},\hat X_{aa},\hat X_{bb},\hat X_{22},\hat X_{ab},\hat X_{ba}, \hat X_{02}\bigg)^T
\end{equation} 
Writing EOMs for these Green functions is not straightforward, since 
$\mathbf{P}^{-1}_{B_1,B_1}$ does not exist. 
Indeed, not every choice of operators yields a matrix which has an inverse. 
For example, commuting operators will  provide a matrix without inverse. 
Two ways to proceed were suggested in Refs.~\onlinecite{TserkovnikovTheorMathPhys81,TserkovnikovTheorMathPhys99,TserkovnikovTheorMathPhys82}:
1.~Complementing the original vector operator with time derivatives of the commuting operators (this approach will resolve the issue, and corresponding formulation will be exact;
however one has to  work with big matrices); and 
2.~Formulations of an approximate practical scheme. 
To proceed we formulate such a scheme,
where instead of single EOM for $\mathbf{G}^{(0)}_{B_1,B_1}(\tau,\tau')$, Eq.(\ref{Astep3}), 
two equations ($s=\pm$) are considered
\begin{align}
\label{Asstep3}
i\frac{\partial}{\partial\tau}\mathbf{G}^{(s,0)}_{B_1,B_1}(\tau,\tau') = 
 \delta(\tau,\tau')\,\mathbf{P}^{(s)}_{B_1,B_1}(\tau) + \mathbf{G}^{(s,0)}_{\dot{B_1},B_1}(\tau,\tau'),
\end{align}
such that $\left[\mathbf{P}^{(s)}_{B_1,B_1}\right]^{-1}$ exists, 
$\mathbf{P}_{B_1,B_1}=\sum_{s=\pm}\mathbf{P}^{(s)}_{B_1,B_1}$, and
$\mathbf{G}^{(0)}_{a,b}(\tau,\tau')=\sum_{s=\pm} \mathbf{G}^{(s,0)}_{a,b}(\tau,\tau')$.
Eqs.~(\ref{Asstep3}) can be solved according to the standard procedure described above,
however each chain has its own definition of the orthogonalized operators,
$\hat A_2^{(s)}\equiv \hat a_2-\mathbf{P}_{a_2A_1}\left[\mathbf{P}_{A_1,A_1}^{(s)}\right]^{-1}\hat A_1$.

In particular, in the simulations below we chose
$
\mathbf{P}^{(s)}_{B_1,B_1}(\tau)=\mathbf{P}_{B_1,B_1}\pm\epsilon\mathbf{I}
$
where $\epsilon$ is a small number and $\mathbf{I}$ is the unit matrix, and
follow the formulation of Section~\ref{methodC}.
In definitions of correlation functions $\mathcal{S}^{(s)}(\tau,\tau')$
(see Eqs.~(\ref{defcalS}) and (\ref{Cfinal}))
for small $\epsilon$ we take approximately 
$\mathbf{G}^{(s,0)}_{a_2,a_2}(\tau,\tau')\approx \mathbf{G}_{a_2,a_2}(\tau,\tau')/2$.
Decoupling system and bath variables in
correlation functions $\mathcal{S}^{(s)}(\tau,\tau')$ this time yields 
Green functions $\mathcal{G}^{(0)}_{A_1,A_1}(\tau,\tau')$,
so that the two sets of equations have to be solved self-consistently.

Figure~\ref{fig4} shows results of calculations which employ the approximate approach
of Section~\ref{methodC} and compares it to  exact results provided by standard NEGF.
Parameters of the calculation are $T=10$~K,  $H^{(M)}_{11}=H^{(M)}_{22}=0.4$~eV,
$H^{(M)}_{12}\equiv t_{12}=0.2$~eV, $\Gamma^L_{11}=\Gamma^R_{22}=0.1$~eV
and $\Gamma^K_{12}=\Gamma^K_{21}=0$ (here $K=L,R$ and
$\Gamma^K_{m_1m_2}\equiv 2\pi\sum_{k\in K} V_{m_1k}V_{km_2}\delta(E-\varepsilon_k)$).
Fermi energy is chosen as the origin, $E_F=0$, and bias $V_{sd}$ is applied on the left,
$\mu_L=E_F+V_{sd}$ and $\mu_R=E_F$. Calculation is performed on an energy grid
spanning the region from $-3$ to $5$~eV with step $10^{-4}$~eV. 
Tolerance for convergence in populations and coherences is $10^{-4}$.
One sees that probabilities of the many-body eigenstates of the system
($P_0=\langle\hat X_{00}\rangle$, $P_a=\langle\hat X_{aa}\rangle$,
$P_b=\langle\hat X_{bb}\rangle$, $P_2=\langle\hat X_{22}\rangle$),
coherences ($\langle X_{ab}\rangle$), and current-voltage characteristics 
of the junctions are accurately reproduced by the approximate procedure 
of Section~\ref{methodC}. Note that standard Redfield QME procedure,
as an approach treating system-bath coupling to finite (second) order,
misses broadening induced by hybridization.

\begin{figure}[htbp]
\centering\includegraphics[width=\linewidth]{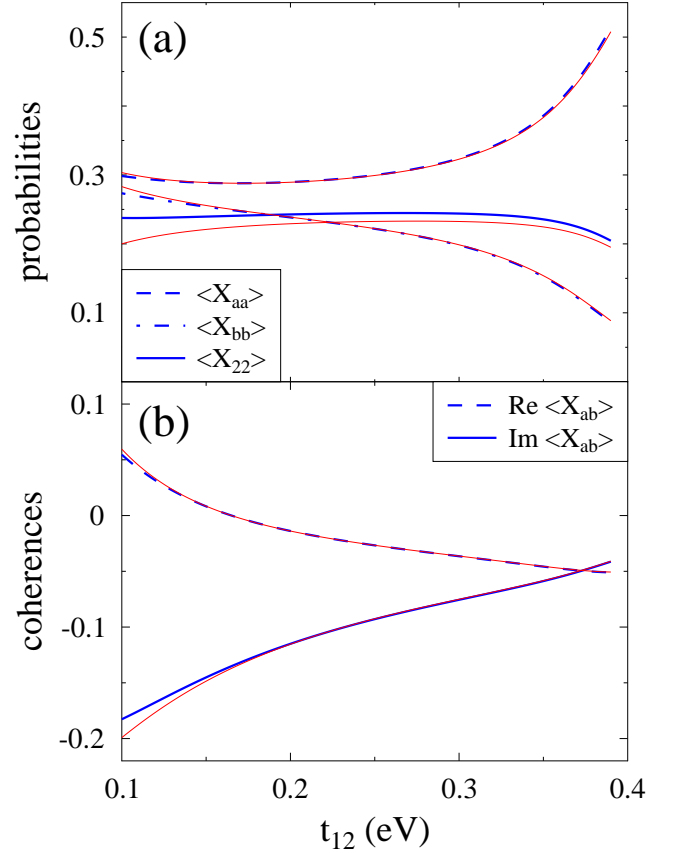}
\caption{\label{fig5}
(Color online) Two-level system (TLS). 
Simulations are performed within the NEGF (solid line; red)
and projection operator scheme of Section~\ref{methodC} 
(solid, dashed, and dash-dotted lines; blue). Shown are (a) probabilities and (b) coherences
of the many-body eigenstates of the TLS vs. electron hopping parameter $t_{12}$
at bias $V_{sd}=0.8$~V. Other parameters are as in Fig.~\ref{fig4}.
}
\end{figure}

Figure~\ref{fig5} shows probabilities and coherences of the many-body states as functions 
of the inter-level electron hopping parameter $t_{12}\equiv H^{(M)}_{12}$.
Calculation is performed at $V_{sd}=0.8$~V, other parameters are as in Fig.~\ref{fig4}.
One sees that the approximate scheme of Section~\ref{methodC} reproduces 
exact results pretty well while the description in terms of non-equilibrium atomic limit
is meaningful, i.e. $t>\Gamma$. 

\section{Conclusions}\label{conclusions}
Progress in experimental techniques at the nanoscale lead to the appearance of new
branches of research. In particular, molecular optoelectronics, nanoplasmonics, and 
spintronics are examples of recent developments in the field of molecular electronics.
These developments provide a challenge for proper theoretical description of 
measurements in strongly interacting open systems far from equilibrium. 
In particular,  development of theoretical methods capable of simulating transport
in molecular junctions in the language of the many-body states of an isolated
system - the nonequilibrium atomic limit - is necessary for
adequate description in such experimentally observed situations as breakdown 
of Born-Oppenheimer approximation, strong electron-vibration interaction 
(polaron formation), avoided crossing (polariton formation) due to strong molecule-plasmon 
interaction, and excitations in single molecule magnetic junctions.

A straightforward general methodology applicable in such situation is provided by the equation-of-motion technique. 
EOM formulated on the Keldysh contour has been used in the literature to describe
transport in junctions by us\cite{GalperinNitzanRatnerPRB06,GalperinNitzanRatnerPRB07} and others.\cite{BulkaKostyrkoPRB04,KostyrkoBulkaPRB05,LevyRabaniJCP13,LevyRabaniJPCM13}
Two main weaknesses of the method are: 1.~The necessity to make uncontrollable approximations in truncating the (in general infinite) chain of EOMs and
2.~Loss of symmetry properties (proper commutation relations) as a result of
decoupling.

Here we introduce a non-equilibrium version of the EOM approach which is 
formulated with the help of projection operators (similar to quantum master equation
formulations), and choice of irreducible Green functions in the EOM chain
(similar to the choice of irreducible diagrams in the standard quantum field theory
methods). The approach, originally developed by Tserkovnikov\cite{TserkovnikovTheorMathPhys81,TserkovnikovTheorMathPhys99}
for description of systems at equilibrium, is reformulated to be applicable to
(in general) time-dependent non-equilibrium situations (Section~\ref{methodA}). 
We derive an explicit canonical form for the chain of EOMs (see Section~\ref{methodB}), 
thus resolving concerns\cite{PedersenPRB09} related to a way of proper construction 
for a  non-equilibrium atomic limit. After a general formulation we discuss application 
of the procedure to transport in junctions. A simple practical approach is proposed
in Section~\ref{methodC}.  
 
This PO-NEGF EOM (PO meaning Projection Operator) formulation is illustrated 
within simple model calculations. In particular, we consider models of
multi-level non-interacting system, quantum dots in the Coulomb blockade and Kondo regimes, and a two-level system treated with Hubbard operators. The latter was the focus
of Ref.~\onlinecite{PedersenPRB09}, when concerns about formulations of
a non-equilibrium atomic limit were raised. 
Results of numerical simulations are compared
to other approximate techniques, and (when available) to exact results.
We show that proper symmetrization,  as discussed
in Refs.~\onlinecite{BonchBruevich_Tyablikov_1962,LevyRabaniJCP13,LevyRabaniJPCM13}, 
is built into the scheme. We also show explicitly that projection (orthogonalization) is crucial 
for proper description of a system coupled to the bath. The latter is missed within
the first Hubbard approximation, as discussed in Refs.~\onlinecite{GalperinNitzanRatnerPRB08,PedersenPRB09} in connection with
the Hubbard non-equilibrium Green function formulation, 
and may be the reason for inconsistencies within the approximation. 

In summary, the presented approach is a step forward in development of NEGF-EOM
approaches. While the proposed truncation procedure 
(similar to the standard EOM methods)
cannot formally guarantee that the resulting approximation will be conserving 
in the sense of being derived from a Luttinger-Ward functional, it introduces two important features:
1. Controllability of the decoupling scheme and 
2. Canonical form of the resulting EOM chain.
The former is due to orthogonalization, Eqs.~(\ref{defAn}), (\ref{defAnd}), and (\ref{defirrG}), and is discussed in detail in Ref.~\onlinecite{TserkovnikovTheorMathPhys99}.
The latter, Eq.(\ref{EOMcqpG}), is a new result, which guarantees Hermiticity
and positive definiteness of the resulting density matrix. 
Moreover, it can be viewed as a way to introduce quasiparticles in the case of strong interactions localized on the molecule. 
Note that the quasiparticles are not molecular (or Kohn-Sham) orbitals of the isolated molecule. Application of the approach to problems, where standard NEGF methodology is not feasible, is the goal of future research.

\begin{acknowledgments}
M.G. gratefully acknowledges support by
the Department of Energy (Early Career Award, DE-SC0006422). 
M.R. thanks the National Science Foundation for support (CHE-1058896).
\end{acknowledgments}
\appendix
\section{Positivity of spectral weights for Fermi-type excitation at finite temperature}\label{app1}
Here we prove that for vector operators 
$A_n(\tau)=(\hat X_{1}(\tau) ,  \ldots, \hat X_{M}(\tau))^T$
($\{\hat X_i(\tau)\}$ are Hubbard operators of Fermi type) 
the spectral weight $\mathbf{P}_{A_n, A_n}(\tau)$, Eq.(\ref{defP}, 
is a positive definite matrix at finite temperatures.
The proof follows from observation that the spectral weight is the sum of two Gram 
matrices, the positive definite character of physical density matrix at finite temperature, 
and properties of the trace.\footnote{In our consideration we follow chapter 1 of 
Ref.~\onlinecite{Bhatia_2007}}

Let $A=(\hat X_1,\dots ,\hat X_M)$ be a vector operator such that $\hat X_i$ is 
an operator of Fermi type for every $i$, then 
\begin{equation}
\mathbf{P}_{A, A}= \mathbf{M}_1+ \mathbf{M}_2
\end{equation}
where
\begin{equation}
\mathbf{M}_1= 
  \begin{bmatrix}
    \langle \hat X_1 \hat X_1^\dagger \rangle & \ldots & \langle \hat X_1 \hat X_M^\dagger \rangle\\
    \vdots&\ddots&\vdots\\
    \langle \hat X_M \hat X_1^\dagger \rangle & \ldots & \langle \hat X_n \hat X_M^\dagger \rangle\\ 
  \end{bmatrix},
\end{equation}
with 
$\langle \hat X_i \hat X_j^\dagger \rangle = \rm{Tr}\{ \hat \rho\, \hat X_i\, \hat X_j^\dagger \}$. 
Similarly, matrix $\mathbf{M}_2$ is composed by 
$\langle\hat X_j^\dagger\hat X_i\rangle$ elements.

For any nonzero $M$-dimensional complex column vector 
${\rm y}=(y_1, \dots, y_M)^T$ we can write 
\begin{equation}
  {\rm y}^\dagger \cdot \mathbf{M}_1 \cdot {\rm y} = 
  \sum_{i,j}^M y_i^{*} \langle \hat X_i \hat X_j^\dagger \rangle y_j
  \equiv \left \langle \hat C\,\hat C^\dagger \right\rangle
\end{equation}
where $\hat C\equiv \sum_{i}^M y_i^{*} \hat X_i$ and we have used that trace is a linear function.  

Since the operators $\hat X_i$ are linearly independent, $\hat C$  is not the zero matrix
in any basis for any non-trivial choice of $\rm y$. 
Let $\rm{c}_i$ be the $i$-th column vector in $\mathbf{C}$. 
By invariance of the trace under cyclic permutations we get 
\begin{equation}
\left \langle \hat C\,\hat C^\dagger \right\rangle
=  \rm{Tr}\{\hat C^\dagger \rho \hat C \} 
= \sum_i \rm{c}_i^\dagger \cdot \mathbf{\rho} \cdot \rm{c}_i 
\end{equation}
Sum on the right is strictly positive as at least one vector $ \rm{c}_i$ is nonzero 
and the density matrix $\mathbf{\rho}$ is positive definite. This proves that
matrix $\mathbf{M}_1$ is positive definite. 

Similarly we can find that 
$\mathbf{M}_2$ is also positive definite matrix. Thus, for any non-zero complex
vector $\rm{y}$
\begin{equation}
{\rm y}^\dagger \cdot \mathbf{P} \cdot {\rm y}
= {\rm y}^\dagger \cdot\left( \mathbf{M}_1+\mathbf{M}_2\right) \cdot {\rm y} > 0
\end{equation}
which completes the proof.

\section{Derivation of Eqs.~(\ref{EOMirrGleft}) and (\ref{EOMirrGright})}\label{appA}
In the derivation below we follow Ref.~\onlinecite{TserkovnikovTheorMathPhys99}.
We derive Eq.(\ref{EOMirrGleft}); derivation of Eq.(\ref{EOMirrGright}) proceeds along 
the same lines. 
 
Staring point is the definition of the non-equilibrium Green function for two arbitrary vector operators
$\hat a$ and $\hat b$ 
\begin{equation}
\label{AdefG}
 \mathbf{G}_{a,b}(\tau,\tau')\equiv -i\langle T_c\,\hat a(\tau)\,\hat b^\dagger(\tau')\rangle
\end{equation}
Its left side EOM can be written as
\begin{equation}
\label{AEOM}
 i\frac{\partial}{\partial\tau}\mathbf{G}_{a,b}(\tau,\tau') = 
 \delta(\tau,\tau')\,\mathbf{P}_{a,b}(\tau) + \mathbf{G}_{\dot{a},b}(\tau,\tau')
\end{equation}
where $\hat{\dot{a}}$ is the time derivative of the operator $\hat{a}$ (in the Heisenberg picture) and $\mathbf{G}_{\dot{a},b}(\tau,\tau')$ is the Green function defined 
by Eq.(\ref{AdefG}) with $\hat a\to \hat{\dot{a}}$. In particular, for orthogonalized operators,
Eqs.~(\ref{defAn}) and (\ref{defAnd}), this becomes
\begin{equation}
\label{AEOMorth}
 i\frac{\partial}{\partial\tau}\mathbf{G}_{A_n,A_m}(\tau,\tau') = 
 \mathbf{G}_{\dot{A}_n,A_m}(\tau,\tau') \qquad (m<n)
\end{equation}
since $\mathbf{P}_{A_n,A_m}(\tau)=0$ for $n\neq m$ by construction.
Utilizing (\ref{AEOM}) and (\ref{AEOMorth}) in the definition of irreducible Green functions,
Eq.(\ref{defirrG}), leads to
\begin{equation}
 \label{Astep3}
 i\frac{\partial}{\partial\tau}\mathbf{G}^{(n-1)}_{A_n,A_n}(\tau,\tau') = 
 \delta(\tau,\tau')\,\mathbf{P}_{A_n,A_n}(\tau) 
 + \mathbf{G}^{(n-1)}_{\dot{A}_n,A_n}(\tau,\tau')
\end{equation}
To find an expression for the last term in the right side of Eq.(\ref{Astep3})
we consider irreducible Green function $\mathbf{G}^{(n)}_{a,\dot{A}_n}(\tau,\tau')$
($\hat a$ is an arbitrary vector operator), which by definition, Eq.(\ref{defirrG}),
satisfies
\begin{align}
 \label{Aint1}
 &\mathbf{G}^{(n)}_{a,\dot{A}_n}(\tau,\tau')= 
 \mathbf{G}^{(n-1)}_{a,\dot{A}_n}(\tau,\tau') 
 \\ &
 - \int_cd\tau_1\int_cd\tau_2\mathbf{G}^{(n-1)}_{a,A_n}(\tau,\tau_1)\,
 \hat{\mathbf{G}}^{-1\,(n-1)}_{A_n,A_n}(\tau_1,\tau_2)\,
 \mathbf{G}^{(n-1)}_{A_n,\dot{A}_n}(\tau_2,\tau')
 \nonumber
\end{align}
Utilizing the right side analog of Eq.(\ref{AEOM}) for the first and last Green functions
in the right side of Eq.(\ref{Aint1}) we get
\begin{align}
\label{Aint2}
&\mathbf{G}^{(n)}_{a,\dot{A}_n}(\tau,\tau')= 
-\delta(\tau,\tau')\,\mathbf{P}_{a,A_n}(\tau)
 \\ &
 + \int_c d\tau_1\mathbf{G}^{(n-1)}_{a,A_n}(\tau,\tau_1)\,
 \hat{\mathbf{G}}^{-1\,(n-1)}_{A_n,A_n}(\tau_1,\tau')\,
 \mathbf{P}_{A_n,A_n}(\tau')
\nonumber
\end{align}
If $\mathbf{P}^{-1}_{A_n,A_n}(\tau)$ exists,\footnote{Here we assume this to be always true. In Section~\ref{numresD} below Eq.(\ref{TLSB1}) we shortly discuss a situation when this is not the case.} 
then Eq.(\ref{Aint2}) can be rewritten in the form
\begin{align}
 \label{Astep4}
 \mathbf{G}_{a,A_n}^{(n-1)}(\tau,\tau') =&
 \int_c d\tau_1\left[\delta(\tau,\tau_1)\,\mathbf{P}_{a,A_n}(\tau)+
 \mathbf{G}^{(n)}_{a,\dot{A}_n}(\tau,\tau_1)\right]\,
 \nonumber \\ &\times
 \mathbf{P}^{-1}_{A_n,A_n}(\tau_1)\,\mathbf{G}^{(n-1)}_{A_n,A_n}(\tau_1,\tau')
\end{align}
Using (\ref{Astep4}) with $\hat a\equiv\hat{\dot{A}}_n$ in (\ref{Astep3})
leads to
\begin{align}
\label{Astep7}
&i\frac{\partial}{\partial\tau}\mathbf{G}^{(n-1)}_{A_n,A_n}(\tau,\tau') =
\delta(\tau,\tau')\,\mathbf{P}_{A_n,A_n}(\tau)
\nonumber \\ &
+\int_c d\tau_1\left[\delta(\tau,\tau_1)\,\mathbf{P}_{\dot{A}_n,A_n}(\tau)
+\mathbf{G}^{(n)}_{\dot{A}_n,\dot{A}_n}(\tau,\tau_1)\right]\,
\\ &\qquad\qquad\times
 \mathbf{P}^{-1}_{A_n,A_n}(\tau_1)\,\mathbf{G}^{(n-1)}_{A_n,A_n}(\tau_1,\tau')
 \nonumber
\end{align} 

Employing vector operator EOMs, Eqs.~(\ref{EOMan}) and (\ref{EOMand}),
in definitions of orthogonalized operators, Eqs.~(\ref{defAn}) and (\ref{defAnd}),
one gets EOMS for the orthogonalized vector operators 
\begin{align}
 \label{AEOMAn}
i\frac{\partial}{\partial\tau}\hat A_n(\tau) =&
\mathbf{\omega}_n(\tau) \hat A_n(\tau) +\mathbf{\nu}_{n,n+1}(\tau)\hat A_{n+1}(\tau)
\nonumber \\
+& \mathbf{\lambda}_{n,n-1}(\tau)\hat A_{n-1}(\tau)
\\
\label{AEOMAnd}
-i\frac{\partial}{\partial\tau'}\hat A_n^\dagger(\tau') =&
\hat A_n^\dagger(\tau')\bar{\mathbf{\omega}}_n(\tau') + 
\hat A_{n+1}^\dagger(\tau')\mathbf{\nu}_{n+1,n}(\tau')
\nonumber\\ +&
\hat A_{n-1}^\dagger(\tau') \mathbf{\lambda}_{n-1,n}(\tau')
\end{align}
where $\mathbf{\omega}_n(\tau)$ is defined in Eq.(\ref{defomega})
($\bar{\mathbf{\omega}}_n=\mathbf{\omega}_n^\dagger$),
$\mathbf{\nu}_{n,n+1}(\tau)$ is defined in Eq.(\ref{EOMan})
($\mathbf{\nu}_{n+1,n}=\mathbf{\nu}_{n,n+1}^\dagger$),
\begin{equation}
\label{Adeflambda}
\mathbf{\lambda}_{n,n-1}(\tau)\equiv\mathbf{P}_{A_n,A_n}(\tau)\,
\mathbf{\nu}_{n,n-1}(\tau)\,\mathbf{P}^{-1}_{A_{n-1},A_{n-1}}(\tau)
\end{equation}
and $\mathbf{\lambda}_{n-1,n}=\mathbf{\lambda}_{n,n-1}^\dagger$.

Finally, substituting (\ref{AEOMAn}) and (\ref{AEOMAnd}) into (\ref{Astep7}),
and utilizing properties of orthogonalized operators 
\begin{equation}
\label{ApropP}
 \mathbf{P}_{A_n,A_m}(\tau)=\delta_{n,m}\,\mathbf{P}_{A_n,A_n}(\tau)
\end{equation}
and irreducible Green functions
\begin{equation}
 \label{ApropG}
 \mathbf{G}^{(n)}_{A_m,A_k}(\tau,\tau') = 0 \qquad
 \mbox{for}\  m\leq n\  \mbox{or}\  k\leq n
\end{equation}
leads to
\begin{align}
 \label{Afinal}
& \left[i\frac{\partial}{\partial\tau}-\mathbf{\omega}_n(\tau)\right]
 \mathbf{G}_{A_n,A_n}^{(n-1)}(\tau,\tau')=
\delta(\tau,\tau')\,\mathbf{P}_{A_n,A_n}(\tau)
\nonumber\\ &
+\int_cd\tau_1 \mathbf{\nu}_{n,n+1}(\tau)\,\mathbf{G}^{(n)}_{A_{n+1},A_{n+1}}(\tau,\tau_1)\,
\mathbf{\nu}_{n+1,n}(\tau_1)\,
\\ &\qquad\qquad\times
\mathbf{P}_{A_n,A_n}^{-1}(\tau_1)\,
\mathbf{G}^{(n-1)}_{A_n,A_n}(\tau_1,\tau')
\nonumber
\end{align}
This completes derivation of Eq.(\ref{EOMirrGleft}).

\section{Derivation of Eq.(\ref{defcqpW})}\label{appB}
Taking the on-the-contour derivative with respect to variable $\tau$ 
of the canonical Green function, Eq.(\ref{defcqpG}), and employing
the EOM for the irreducible Green function, Eq.(\ref{EOMirrGleft}), results in EOM (\ref{EOMcqpG})
with the asymmetric free evolution term of the form
\begin{align}
 \label{Bstart}
 \mathcal{W}_n(\tau) =&
 i\frac{\partial}{\partial\tau}\left[\mathbf{P}_{A_n,A_n}^{-1/2}(\tau)\right]\,
 \mathbf{P}_{A_n,A_n}^{1/2}(\tau) 
 \\ +& \mathbf{P}_{A_n,A_n}^{-1/2}(\tau)\,
 \mathbf{\omega}_n(\tau)\,\mathbf{P}_{A_n,A_n}^{1/2}(\tau)
 \nonumber
\end{align}
Starting point of our consideration is the definition of the normalized free evolution
matrix $\mathbf{\omega}_n(\tau)$, Eq.(\ref{defomega}). Our goal is to rewrite 
the second and third terms in the right side of Eq.(\ref{defomega}) is such a way,
that the canonical free evolution term, Eq.(\ref{Bstart}), will be represented in a
clear symmetrized form. 

First, consider the second term in the right side of Eq.(\ref{defomega}). 
Using the definition of the scalar product,
Eq.(\ref{defP}), and vector-operator EOM, Eq.(\ref{EOMan}), we can write
(to save space below we drop dependence on contour variable $\tau$; such dependence 
is assumed for every element in the following expressions)
\begin{align}
 \label{Bint1}
 &\mathbf{\nu}_{n,n+1}\,\mathbf{P}_{a_{n+1},A_n}\,\mathbf{P}^{-1}_{A_n,A_n} =
 \left\langle\left[\mathbf{\nu}_{n,n+1}\hat a_{n+1};\hat A_n^\dagger\right]_{\pm}\right\rangle
 \,\mathbf{P}^{-1}_{A_n,A_n}
 \\ &=
 \left\langle\left[i\hat{\dot{a}}_n;\hat A_n^\dagger\right]_\pm\right\rangle
 \,\mathbf{P}^{-1}_{A_n,A_n}
 -\mathbf{\varepsilon}_n\left\langle\left[\hat a_n;\hat A_n^\dagger\right]_\pm\right\rangle
 \,\mathbf{P}^{-1}_{A_n,A_n}
 \nonumber
\end{align}
From the definition of orthogonalized operators, Eq.(\ref{defAn}),
it is easy to see that $\mathbf{P}_{a_n,A_n}=\mathbf{P}_{A_n,A_n}$.
Thus the last term in Eq.(\ref{Bint1}) becomes $\mathbf{\varepsilon}_n$.
The term preceding it can be evaluated as
\begin{equation}
  \bigg(
  i\frac{\partial}{\partial\tau}\left[\mathbf{P_{A_n,A_n}}\right]
 +\left\langle\left[\hat a_n;-i\hat{\dot{A}}_n^\dagger\right]_\pm\right\rangle
 \bigg)\,\mathbf{P}^{-1}_{A_n,A_n}
\end{equation}
Employing in the latter expression EOM (\ref{AEOMAnd}), and combining the results 
in (\ref{Bint1}) leads to 
\begin{align}
\label{Bfin1}
&\mathbf{\nu}_{n,n+1}\,\mathbf{P}_{a_{n+1},A_n}\,\mathbf{P}^{-1}_{A_n,A_n} =
\bigg(i\frac{\partial}{\partial\tau}\left[\mathbf{P_{A_n,A_n}}\right]
\\ &
+\mathbf{P}_{A_n,A_n}\,\bar{\mathbf{\omega}}_n
+\mathbf{P}_{a_n,A_{n-1}}\,\mathbf{\lambda}_{n-1,n}
\bigg)\,\mathbf{P}^{-1}_{A_n,A_n}
-\mathbf{\varepsilon}_n
\nonumber
\end{align}
 
Now let us consider the third term in the right of Eq.(\ref{defomega}). We rewrite it as
\begin{align}
&\mathbf{P}_{a_n,A_{n-1}}\,\mathbf{P}^{-1}_{A_{n-1},A_{n-1}}\,\mathbf{\nu}_{n-1,n}
\nonumber \\ &
\equiv \mathbf{P}_{a_n,A_{n-1}}\,\mathbf{P}^{-1}_{A_{n-1},A_{n-1}}\,\mathbf{\nu}_{n-1,n}
\mathbf{P}_{A_n,A_n}\,\mathbf{P}^{-1}_{A_n,A_n}
\\ & =
\mathbf{P}_{a_n,A_{n-1}}\,\mathbf{P}^{-1}_{A_{n-1},A_{n-1}}\,
\left\langle\left[\nu_{n-1,n}\hat A_n;\hat A_n^\dagger\right]_\pm\right\rangle
\,\mathbf{P}^{-1}_{A_n,A_n}
\nonumber
\end{align}
Utilizing the EOM (\ref{AEOMAn}) and performing evaluations similar to those which
yield Eq.(\ref{Bfin1}) leads to
\begin{equation}
 \label{Bfin2}
 \mathbf{P}_{a_n,A_{n-1}}\,\mathbf{P}^{-1}_{A_{n-1},A_{n-1}}\,\mathbf{\nu}_{n-1,n}
 = \mathbf{P}_{a_n,A_{n-1}}\,\mathbf{\lambda}_{n-1,n}\,\mathbf{P}^{-1}_{A_n,A_n}
\end{equation}

Substituting (\ref{Bfin1}) and (\ref{Bfin2}) into (\ref{defomega}) yields
\begin{align}
\label{Bfin}
\mathbf{\omega}_n(\tau) =& i\frac{\partial}{\partial\tau}\left[\mathbf{P}_{A_n,A_n}(\tau)\right]\,
\mathbf{P}_{A_n,A_n}^{-1}(\tau)
\\ +& \mathbf{P}_{A_n,A_n}(\tau)\,\bar{\mathbf{\omega}}_n(\tau)\,
\mathbf{P}_{A_n,A_n}^{-1}(\tau)
\nonumber
\end{align}
Finally, using (\ref{Bfin}) in (\ref{Bstart}), and employing the matrix relation
\begin{equation}
\mathbf{M}^{1/2}(x)\,\frac{d}{dx}\left[\mathbf{M}^{-1/2}(x)\right] =
- \frac{d}{dx}\left[\mathbf{M}^{1/2}(x)\right]\,\mathbf{M}^{-1/2}(x)
\end{equation}
leads to Eq.(\ref{defcqpW}).

\section{Derivation of Eq.(\ref{EOMcqpS})}\label{appC}
We start from definition of the irreducible Green function, Eq.(\ref{defirrG}),
which for $\hat a=\hat b=\hat A_2$ is
\begin{align}
\label{Cstart}
 & \mathbf{G}^{(1)}_{A_2,A_2}(\tau,\tau') = \mathbf{G}^{(0)}_{A_2,A_2}(\tau,\tau')
 \\ &
 -\int_c d\tau_1\int_c d\tau_2\, \mathbf{G}^{(0)}_{A_2,A_1}(\tau,\tau_1)\,
 \mathbf{G}^{-1\,(0)}_{A_1,A_1}(\tau_1,\tau_2)\,\mathbf{G}^{(0)}_{A_1,A_2}(\tau_2,\tau')
 \nonumber
\end{align}
Using (\ref{Astep4}) with $\hat a=\hat A_2$ and $n=1$ leads to
\begin{align}
 \label{Cstep1}
 \mathbf{G}^{(0)}_{A_2,A_1}(\tau,\tau') =& \int_c d\tau_1\,
 \mathbf{G}^{(1)}_{A_2,A_2}(\tau,\tau_1)\,\mathbf{\nu}_{2,1}(\tau_1)\,
 \\ &\times
 \mathbf{P}^{-1}_{A_1,A_1}(\tau_1)\,\mathbf{G}^{(0)}_{A_1,A_1}(\tau_1,\tau')
 \nonumber
\end{align}
where we used the EOM (\ref{AEOMAn}) and properties (\ref{ApropP}) and (\ref{ApropG}).
Consideration similar to that which leads from Eq.(\ref{Aint1}) to (\ref{Astep4})
performed for Green function $\mathbf{G}^{(n)}_{\dot{A}_n,b}(\tau,\tau')$,
after taking $\hat b=\hat A_2$ leads to
\begin{align}
\label{Cstep2}
 \mathbf{G}^{(0)}_{A_1,A_2}(\tau,\tau') =& \int_c d\tau_1\,
 \mathbf{G}^{(0)}_{A_1,A_1}(\tau,\tau_1)\,\mathbf{P}^{-1}_{A_1,A_1}(\tau_1)\,
 \\ &\times
 \mathbf{\nu}_{1,2}(\tau_1)\,\mathbf{G}^{(1)}_{A_2,A_2}(\tau_1,\tau')
 \nonumber
\end{align}
Substituting (\ref{Cstep1}) and (\ref{Cstep2}) into (\ref{Cstart}) yields
\begin{align}
 \label{Cintermediate}
 &\mathbf{G}^{(1)}_{A_2,A_2}(\tau,\tau') = \mathbf{G}^{(0)}_{A_2,A_2}(\tau,\tau')
 \nonumber \\ &
 -\int_cd\tau_1\int_cd\tau_2 \mathbf{G}^{(1)}_{A_2,A_2}(\tau,\tau_1)\,
 \mathbf{\nu}_{2,1}\,\mathbf{P}^{-1}_{A_1,A_1}(\tau_1)\,
 \\ &\quad\times
 \mathbf{G}^{(0)}_{A_1,A_1}(\tau_1,\tau_2)\,\mathbf{P}^{-1}_{A_1,A_1}(\tau_2)\,
 \mathbf{\nu}_{1,2}\,\mathbf{G}^{(1)}_{A_2,A_2}(\tau_2,\tau')
 \nonumber
\end{align}
Finally, substituting definitions of orthogonalized operators, Eqs.~(\ref{defAn}) and
(\ref{defAnd}), into the irreducible Green function $\mathbf{G}^{(0)}_{A_2,A_2}(\tau,\tau')$,
and performing transformations similar to those leading to Eqs.~(\ref{Cstep1})
and (\ref{Cstep2}) on Green functions $\mathbf{G}^{(0)}_{a_2,A_1}(\tau,\tau')$
and $\mathbf{G}^{(0)}_{A_1,a_2}(\tau,\tau')$ results in
\begin{widetext}
\begin{align}
 \label{Cfinal}
 \mathbf{G}^{(1)}_{A_2,A_2}(\tau,\tau') =& \mathbf{G}^{(0)}_{A_2,A_2}(\tau,\tau')
-\mathbf{P}_{a_2,A_1}(\tau)\,\mathbf{P}^{-1}_{A_1,A_1}(\tau)\,
 \mathbf{G}^{(0)}_{A_1,A_1}(\tau,\tau')\,
 \mathbf{P}^{-1}_{A_1,A_1}(\tau')\,\mathbf{P}_{A_1,a_2}(\tau')
 \nonumber \\ -&
 \int_c d\tau_1\bigg(
  \mathbf{P}_{a_2,A_1}(\tau)\,\mathbf{P}^{-1}_{A_1,A_1}(\tau)\,
  \mathbf{G}^{(0)}_{A_1,A_1}(\tau,\tau_1)\,
  \mathbf{P}^{-1}_{A_1,A_1}(\tau_1)\,
  \mathbf{\nu}_{1,2}(\tau_1)\,\mathbf{G}^{(1)}_{A_2,A_2}(\tau_1,\tau')
  \\ & \qquad\  +
  \mathbf{G}^{(1)}_{A_2,A_2}(\tau,\tau_1)\,\mathbf{\nu}_{2,1}(\tau_1)\,
 \mathbf{P}^{-1}_{A_1,A_1}(\tau)\,
 \mathbf{G}^{(0)}_{A_1,A_1}(\tau_1,\tau')\,
 \mathbf{P}^{-1}_{A_1,A_1}(\tau')\,\mathbf{P}_{A_1,a_2}(\tau')
 \bigg)
 \nonumber \\ -&
 \int_c d\tau_1\int_c d\tau_2\, \mathbf{G}^{(1)}_{A_2,A_2}(\tau,\tau_1)\,
 \mathbf{\nu}_{2,1}(\tau_1)\,\mathbf{P}^{-1}_{A_1,A_1}(\tau_1)\,
 \mathbf{G}^{(0)}_{A_1,A_1}(\tau_1,\tau_2)\,
 \mathbf{P}^{-1}_{A_1,A_1}(\tau_2)\,\mathbf{\nu}_{1,2}(\tau_2)\,
 \mathbf{G}^{(1)}_{A_2,A_2}(\tau_2,\tau')
 \nonumber
\end{align}
\end{widetext}
This completes the derivation. Eq.(\ref{EOMcqpS}) directly follows from (\ref{Cfinal})
when using the definition of the canonical self-energy, Eq.(\ref{defcqpS}).
\section{Hermiticity and positive definiteness of the calculated density matrix}\label{app2}
Here we prove that calculated physical density matrix, which for Fermi type excitations
in the system is given by lesser and greater projections of the physical Green function, 
Eq.(\ref{defG}), taken at equal times is positive definite matrix at finite temperatures.

The proof follows from the Dyson type of EOM for the canonical Green function,
Eq.(\ref{EOMcqpG}), the connection between the physical and the canonical Green functions,
Eq.(\ref{defcqpG}), and the positivity at finite temperature of the spectral weight matrix 
(see Appendix~\ref{app1}). 

First, truncating the infinite EOM chain, Eq.(\ref{EOMcqpG}),
at some step $n$ in the hierarchy by neglecting self-energy $\mathcal{S}^{(n)}$, 
one gets free evolution Dyson equation for the canonical GF $\mathcal{G}^{(n)}$,
projections of which have usual proper relations among themselves.
This in turn means that projections of the self-energy of the previous step, 
Eq.(\ref{defcqpS}), all fulfill all the usual relations. Going this way up in the chain of EOMs
one arrives to the top, $n=1$, Dyson equation, which by its canonical form guarantees 
positivity of $i\mathcal{G}^{>}(t,t)$, $-i\mathcal{G}^{<}(t,t)$, and 
the canonical density matrix.

Second, since canonical an physical GFs are related  by the scaling, Eq.(\ref{defcqpG}), 
defined by positive definite matrices, this implies that matrices
$i\mathbf{G}^{>}(t,t)$ and $-i\mathbf{G}^{<}(t,t)$ are also positive definite.\footnote{This statement is also discussed in chapter 6 of Ref.~\onlinecite{Bhatia_2007} (see discussion below Eq.(6.2) on p.201)}
Indeed,
let ${\rm x}$ be any complex column vector of the same dimension as ${\bold G}_{A_n, A_n}^{<} (t,t)$. Then
\begin{align}
  {\rm x}^\dagger \cdot  {\bold G}_{A_n, A_n}^{<} (t,t) \cdot  {\rm x}=&  {\rm x}^\dagger \cdot {\bold P}_{A_n, A_n}^{1/2} \mathcal{G}_{A_n, A_n}^{<}(t,t) {\bold P}_{A_n, A_n}^{1/2} \cdot {\rm x}
  \nonumber \\
=& {\rm y}^\dagger\cdot \mathcal{G}_{A_n, A_n}^{<} (t,t) \cdot {\rm y}
\end{align}  
with ${\rm y} = {\bold P}_{A_n, A_n}^{1/2} \cdot {\rm x}$. 
Thus ${\rm x}^\dagger \cdot  {\bold G}_{A_n, A_n}^{<} (t,t) \cdot  {\rm x} >0$ 
if and only if $ {\rm y}^\dagger\cdot \mathcal{G}_{A_n, A_n}^{<} (t,t) \cdot {\rm y} >0$ 
as we can write ${\rm x} = {\bold P}_{A_n, A_n}^{-1/2} \cdot {\rm y}$ for arbitrary ${\rm y}$.

Finally, positivity of lesser and greater projection of the physical GF,
$i\mathbf{G}^{>}(t,t)$ and $-i\mathbf{G}^{<}(t,t)$ implies positivity of the 
physical density matrix. This shows consistency of our truncation scheme.

\bibliography{projGF5}

\end{document}